\documentclass[aps,pra,twocolumn,10pt,showpacs]{revtex4-2}
\pdfoutput=1

\usepackage{graphicx}
\usepackage{amsmath,amssymb}
\usepackage{hyperref}
\hypersetup{colorlinks=true,
						linkcolor=red,
						anchorcolor=blue,
						citecolor=blue,
						filecolor=blue,
						menucolor=blue,
						urlcolor=blue
}

\usepackage{color}
\usepackage{braket}
\usepackage[english]{babel}

\usepackage{mathdots}
\usepackage{bbold}

\newcommand*\diff{\mathop{}\!\mathrm{d}}

\begin{document}
\title{Floquet engineering of individual band gaps in an optical lattice using a two-tone drive}
\author{Kilian Sandholzer}
\author{Anne-Sophie Walter}
\author{Joaqu\' in Minguzzi}
\author{Zijie Zhu}
\author{Konrad Viebahn}
\author{Tilman Esslinger}
\affiliation{Institute for Quantum Electronics, ETH Zurich, 8093 Zurich, Switzerland}

\date{\today}

\begin{abstract}
The dynamic engineering of band structures for ultracold atoms in optical lattices represents an innovative approach to understand and explore the fundamental principles of topological matter.
In particular, the folded Floquet spectrum determines the associated band topology via band inversion.
We experimentally and theoretically study two-frequency phase modulation to asymmetrically hybridize the lowest two bands of a one-dimensional lattice.
Using quasi-degenerate perturbation theory in the extended Floquet space we derive an effective two-band model that quantitatively describes our setting.
The energy gaps are experimentally probed via Landau-Zener transitions between Floquet-Bloch bands using an accelerated Bose-Einstein condensate.
Separate and simultaneous control over the closing and reopening of these band gaps is demonstrated.
We find good agreement between experiment and theory, establishing an analytic description for resonant Floquet-Bloch engineering that includes single- and multi-photon couplings, as well as interference effects between several commensurate drives.
\end{abstract}

\maketitle

The quantum states emerging in periodic potentials are based on the properties of the underlying band structure.
Its symmetry induced topology \cite{Schnyder2008,Kitaev2009,Ryu2010,Fu2011,Slager2013,Dong2016,Chiu2016,Roy2017} leads to special types of band insulators which are robust against perturbations conserving the protecting symmetries \cite{Hasan2010,Qi2011,Bernevig2013}.
The experimental realization of such systems is a crucial step to further understand their foundations and study the dynamic properties of the states.
The key for engineering such topological band structures lies in the individual control of degeneracies at band inversion points \cite{Rudner2020}.
Starting from a topologically trivial band structure, the induced band inversion points provide the necessary gap closing transition.
However, it is a challenge to achieve dynamic control of band structures and couplings in experimental realizations \cite{Ando2013,Ando2015,Cooper2019}.

Floquet band engineering introduces a tool to change the band properties and opens up a path to dynamically study the basic mechanisms of topological matter \cite{Rudner2020,Kitagawa2010,Cayssol2013,Gomez-Leon2013,Oka2019}.
It has been studied in photonic systems \cite{Rechtsman2013,Ozawa2019}, in solid state materials \cite{Wang2013,McIver2020}, and ultracold atoms \cite{Cooper2019,Goldman2016a,Weitenberg2021}, the latter providing a possibility to introduce tunable interactions \cite{Tai2017}.
While the usage of bipartite, two-dimensional optical lattices creates tunable band inversion points in form of Dirac points \cite{Tarruell2012}, circular shaking in such a system \cite{Jotzu2014} controls the individual gaps at these points.
Besides the two-dimensional implementations, a fundamental understanding of topological matter can be gained in one-dimensional implementations.
This has been pursued either in bipartite lattices \cite{Atala2013,Nakajima2016} or by using synthetic dimensions \cite{Lin2011,Galitski2013,Livi2016,Kolkowitz2017,Meier2016} and single frequency resonant shaking \cite{Kang2020a}.
In this work we combine a simple, one-dimensional lattice with Floquet engineering using a two-frequency driving scheme to obtain full control on the band inversion points and their gaps.
The induced destructive interference by the two commensurate frequencies depends on quasimomentum.
This allows us to engineer the coupling at individual band inversion points as well as dynamically decouple a full band.
We derive an effective model using quasi-degenerate perturbation theory and probe the bandstructure with ultracold atoms in optical lattices.
\begin{figure*}
    \includegraphics{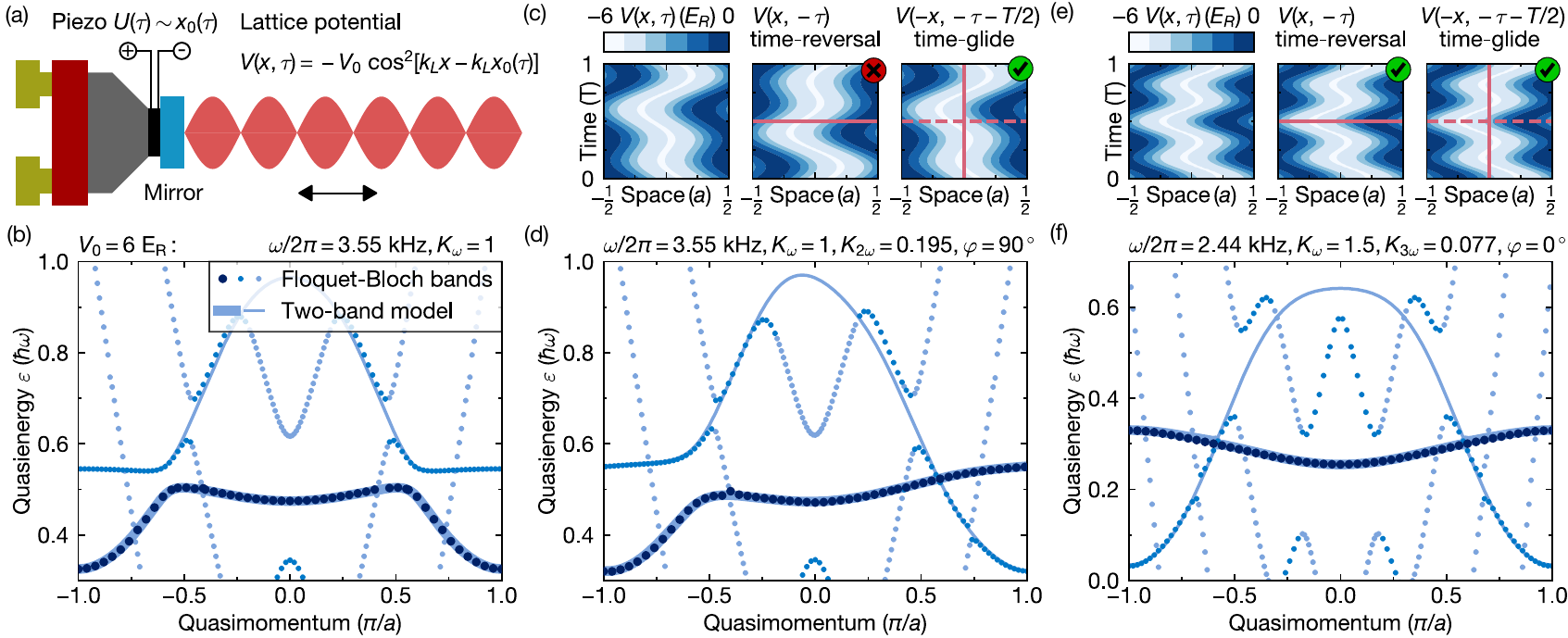}
    \caption{Floquet band engineering with multi-frequency phase modulation. In part (a) the basic experimental setup is sketched: an optical lattice potential $V(x,\tau)$ created by a retro-reflected beam (wavevector $k_L$) is phase modulated by displacing the mirror with a piezo-electric actuator. The position of the mirror $x_0(\tau)$ is proportional to the voltage $U(\tau)$ applied to the piezo-electric device. If the modulation frequency is resonant to the band gap, the $s$- and $p$-band hybridize to an effective Floquet band, shown in the spectrum for the quasienergy $\varepsilon$ (b). For a two-color drive with fundamental and second harmonic ($\omega,2\omega$), there exists a driving waveform at relative phase $\varphi=90^\circ$ for which time-reversal symmetry of the potential $V(x,\tau)$ is broken but time-glide symmetry is conserved (c). The resulting spectrum (d) is asymmetric in quasimomentum and for driving strengths of $K_\omega=1$ and $K_{2\omega}=0.195$, the gap at positive quasimomentum closes. If only fundamental and third harmonic are included in the drive ($\omega,3\omega$), we restore both time-reversal and time-glide symmetry for a relative phase of $\varphi=0^\circ$ (e) which makes it possible to decouple the Floquet $s$-band from higher bands (f). We show the Floquet-Bloch bands with most overlap to the $s$-, $p$- and $d$-bands of a static lattice as dots. These bands are obtained via diagonalization of the evolution operator of one period \cite{Suppl}.  We also show the effective bands (solid lines) of an analytic two-band model derived using quasi-degenerate perturbation theory \cite{Suppl}.
    \label{fig1}}
\end{figure*}

In the initial one-dimensional lattice the lowest bands are energetically well separated.
If the position of the potential is modulated periodically in time, we can use Floquet analysis \cite{Holthaus2015,Eckardt2017} to calculate the quasienergy spectrum for the atoms. 
Since the drive provides energy in multiples $l$ of $\hbar\omega$, we can create band inversion points by choosing the drive frequency resonant to the gap $\Delta(q)$ between $s$- and $p$-band at a specific quasimomentum value $l\hbar\omega = \Delta(q_c)$.

While this coupling can be used transiently to probe \cite{Schori2004,Heinze2011,Flaschner2018} or manipulate \cite{Fujiwara2019,Arnal2019} the state of the static system, we are interested in the effective physics induced by the Floquet band structure.
The direct coupling of $s$- and $p$-band in a one-dimensional lattice using single frequency shaking has been implemented to create hybridized effective bands populated by a Bose-Einstein condensate (BEC) \cite{Parker2013} including the study of interaction effects \cite{Ha2015,Song2021}.
Beyond the usage of the direct coupling mechanism, the understanding of multi-photon resonances \cite{Weinberg2015,Cabrera-Gutierrez2019} enabled the implementation and characterization of quasimomentum dependent couplings \cite{Kang2018,Kang2020a}.
In addition, by extending purely harmonic modulation to polyharmonic or anharmonic driving waveforms it is possible to break time-reversal symmetry which allows for the realization of asymmetric band structures \cite{Struck2012,Grossert2016}.
The combination of time-reversal symmetry breaking and multi-photon resonances has been applied to Fermions in shaken lattices resonant to the interaction \cite{Gorg2019}, near-resonant driving to the $sp$-band gap \cite{Yao2021}, resonant amplitude modulations \cite{Niu2015,Viebahn2021} and mixed schemes \cite{Zhuang2013,Grossert2016}.
We extend this method to phase modulated lattices resonant with the $sp$-band gap to create asymmetric, hybridized bands and control the individual gaps.

The conceptual and experimental setup consists of a retro-reflected laser beam creating the underlying periodic potential for ultracold atoms. 
The single particle spectrum is defined by a one-dimensional lattice Hamiltonian
\begin{equation}
\hat{H}_\text{sp} = \frac{\hat{p}^2}{2M}-V_0 \cos^2\left[k_L \hat{x}-k_L x_0\left(\tau\right)\right].
\label{eqn:Hlab}
\end{equation}
The depth $V_0$ and phase $k_L x_0$ can be controlled externally by varying the intensity of the laser and the position of the retro-reflecting mirror.
A piezo-electric actuator gives precise and fast control \cite{Suppl} on the mirror position defining the phase of the lattice potential (see Fig.~\ref{fig1}a)
\begin{align}
k_L x_0 (\tau) =\ \frac{2 E_{\text{rec}}}{\pi \hbar \omega} \left[K_\omega \cos\left(\omega \tau\right) +  \frac{K_{l\omega} }{l} \cos\left(l\omega\tau+ \varphi\right)\right].
\end{align}
The amplitude is parametrized by the recoil energy $E_{\text{rec}}= \hbar^2 k_L^2/2M$, where $k_L=\pi/a=2\pi/\lambda_L$ is the wave vector of the lattice laser, the angular frequency $\omega$ and the dimensionless driving strengths $K_\omega$, $K_{l\omega}$, with $l \in \left[2,3\right]$.
The driving strength $K_\omega$ is connected to the expansion of the piezo-electric actuator $\Delta L_\omega$ via $K_\omega = \pi^2(\Delta L_\omega/\lambda_L)(\hbar \omega/E_{\text{rec}})$.
For the experiments in this paper we use $^{87}$Rb and a laser wavelength of $\lambda_L=1064\ \mathrm{nm}$ which gives $E_R/h = 2026$~Hz using the mass $M$ of $^{87}$Rb \cite{Suppl}.

The frequency of the periodic forcing is set on resonance to an integer multiple of the band gap between the $s$- and $p$-band of the lattice leading to band inversion in the folded Floquet spectrum.
The periodic forcing induces interband transitions versus quasimomentum, creating avoided crossing in this parameter.
In general, the von Neumann-Wigner non-crossing rule \cite{VonNeumann1929} establishes a gap opening in quasimomentum for single harmonic driving as shown in Fig.~\ref{fig1}(b).
The lowest band and first excited band become hybridized.
We focus our studies on the Floquet band with $s$-band character in the center of the Brillouin zone and $p$-band character at the edge.
The $p$-band part is dressed with an energy quanta from the drive which shifts the minimum of the band from $q=0$ to $q=\pm \pi/a$.
Since the structure of this band is mainly defined by the $s$- and $p$-band, a two-band model is sufficient to capture the dynamics.
The other hybridized bands include major contribution from $d$- and higher bands and multi-band models are necessary to fully describe their physics.

The addition of higher harmonics to the drive allows us to use constructive and destructive interference on the interband couplings and to shape the dispersion of the desired effective band.
The topology of the hybridized bands can be related to the space-time symmetry of the periodic driving potential \cite{Gomez-Leon2013,Morimoto2017,Xu2018}.
In the case of driving the system with the fundamental and second harmonic $l=2$ at a relative phase of $\varphi=90^\circ$ the potential breaks time-reversal symmetry as shown in Fig.~\ref{fig1}(c), leading to a band structure asymmetric in quasimomentum.
However, it preserves time-glide symmetry (space mirror plus half period time translation) which makes it possible to close a single gap at only one half of the Brillouin zone as shown in Fig.~\ref{fig1}(d).
The closing and reopening of a gap in the spectrum signals a possible topological phase transition and therefore constitutes an optimal handle to engineer topological one-dimensional structures \cite{Rudner2020}.
If we choose a fundamental and third harmonic $l=3$ driving with $\varphi=0^\circ$, time-reversal symmetry as well as time-glide symmetry are preserved as depicted in Fig.~\ref{fig1}(e).
The resulting spectrum is symmetric in quasimomentum and the opened gaps are closed simultaneously at both quasimomenta as shown in Fig.~\ref{fig1}(f).
Since the band dispersion is still mainly defined by the $l=1$ Floquet drive this method is well suited to suppress heating to higher bands in a strongly driven lattice \cite{Viebahn2021}.

Single particles in the one-dimensional lattice with two-frequency modulation can be described in the language of spatially and temporally periodic Floquet-Bloch wavefunctions.
Their spectrum [Fig.~\ref{fig1}(b,d,f)] is obtained by diagonalizing the one-period evolution operator \cite{Holthaus2015}.
In addition, we employ quasi-degenerate perturbation theory on the extended Floquet space \cite{Eckardt2015} which has been used for resonant single-frequency driving \cite{Weinberg2015} and is extended to two-frequency schemes in this work. 
This approach allows us to derive a precise effective Hamiltonian model for the real experimental implementation.
The method is equivalent to a high-frequency expansion but the extended space picture in combination with polychromatic driving allows us to intuitively design the driving waveform to construct a specific effective Hamiltonian.
In both approaches, the driving is implemented as a time-dependent gauge field which corresponds to the Hamiltonian of Eq.~\ref{eqn:Hlab} via a basis transformation \cite{Suppl}.

To arrive at the extended space quasienergy operator $\hat{Q}$, we transform the Hamiltonian to the co-moving frame, where it can be represented in a basis of time-dependent Bloch states, i.e.
\begin{align}
\hat{H}(\tau)= & \sum_{q\left(\tau\right),n} \left\{\varepsilon_n \left[q\left(\tau\right)\right] \hat{c}_{q\left(\tau\right),n}^\dagger \hat{c}_{q\left(\tau\right),n} \right.\nonumber\\
& \left. + M\ddot{x}_0(\tau) \sum_{n^\prime} \eta_{n n^\prime}\left[q\left(\tau\right)\right]\hat{c}_{q\left(\tau\right),n}^\dagger \hat{c}_{q\left(\tau\right),n^\prime}\right\}.
\end{align}
In this frame the Hamiltonian separates into the dispersion $\varepsilon_n\left[q\left(\tau\right)\right]$ of a Bloch state in band $n$ for a time dependent quasimomentum $q\left(\tau\right)=q-M\dot{x}_0/\hbar$ and the inter-band coupling element $\eta_{nn^\prime}\left[q\left(\tau\right)\right]$.
The Fourier coefficients of this Hamiltonian $\hat{H}_m$ are the building blocks of the quasienergy operator $\hat{Q}$ expressed in the extended Floquet-Bloch basis $|nqm\rangle\rangle=|nq\rangle e^{im\omega \tau}$.
\begin{align}
\langle\langle n^\prime q^\prime m^\prime | \hat{Q} | n q m \rangle\rangle = & \langle n^\prime q^\prime | \hat{H}_{m-m^\prime}| n q \rangle \nonumber\\
&+\delta_{mm^\prime}\delta_{qq^\prime} \left(m-m^\prime\right) \hbar \omega.
\end{align}
We use a tight-binding approximation to find an expression for the Fourier coefficients for the dispersion $\varepsilon_n$ and the inter-band coupling $\eta_{nn^\prime}$.
While the tight-binding (nearest neighbor) approximation is sufficient to describe the lowest band, higher order terms (longer range tunneling) must be incorporated for the $p$-band.
Equivalently, extended (longer range) interband coupling terms are taken into account in the calculations to accurately model the band hybridization, more details can be found in the Appendix.

\begin{figure}
    \includegraphics{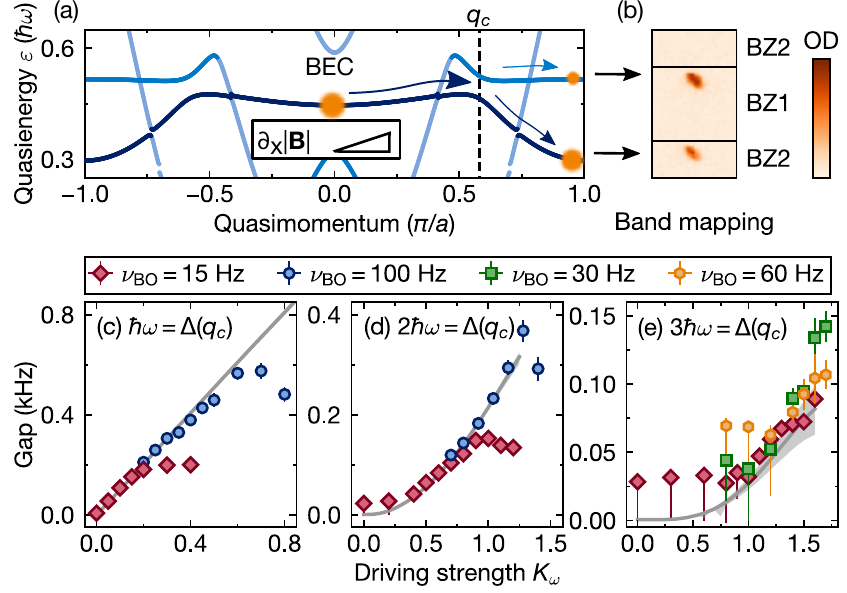}
    \caption{Experimental probe of the Floquet band structure. (a) We load two-dimensional pancakes of a BEC in a one-dimensional shaken lattice (x-direction) and apply a magnetic field gradient $\partial_x |\mathbf{B}|$. The resulting force on the atoms induces Bloch oscillations and atoms will transfer from the $s$-band to the $p$-band at the coupling point $q_c$ for resonant modulation. (b) The clouds in $s$- and $p$-band are separately detected using band mapping. We extract the transferred fraction from the absorption images by fitting a Gaussian to the optical density (OD). In (c-e) we plot the measured gap versus the dimensionless driving strength $K_\omega$ and compare it to a numerical simulation of the Floquet-Bloch spectrum. The gap size is determined with the Landau-Zener formula from the measured transition rates. In panel (c) we probe a one-photon resonance $\hbar\omega=\Delta(q_c)$, for panel (d) a two-photon resonance, $2\hbar\omega=\Delta(q_c)$, and for panel (e) a three-photon resonance, $3\hbar\omega=\Delta(q_c)$. The Bloch oscillation frequency $\nu_\mathrm{BO}$ determines the resolution of the gap measurement, a saturation effect appears in the data when the resolution limit is reached. Error bars on the experimental data include the standard error of four measurements as well as systematic errors due to uncertainties in the calibration of lattice depth, shaking strength and Landau-Zener transition speed. The shaded area for theoretical curves reflects the statistical and systematic error in the lattice depth.  \label{fig2}}	
\end{figure}
To probe the structure, we load a BEC of $^{87}$Rb atoms into the $s$-band of a one-dimensional lattice, creating pancakes of two-dimensional condensates.
After ramping up the shaking waveform we use a magnetic field gradient to induce Bloch oscillations as a probe for the Floquet-Bloch spectrum \cite{Fujiwara2019}.
The atoms sweep through different quasimomentum states.
At the coupling point, they are partially transferred to the $p$-band in a Landau-Zener process.
Subsequent band mapping of the cloud reveals the transferred fraction, shown in Figure~\ref{fig2}(a), and provides a measure for the gap energy using the Landau-Zener formula \cite{Zener1932,Landau1932}.
The sensitivity of this method is limited by how slow the Bloch oscillations can be done.
In our system, the main decoherence effect of the hybridized band is caused by dipole oscillations through the underlying harmonic confinement.
Since this confinement is needed to support the atoms against gravity we use magnetic levitation to minimize the trapping frequency to $f_x^\text{trap}=7.4(3)\ \text{Hz}$.
We achieve reliable results for Bloch oscillation frequencies down to 15 Hz.

We test the method on three different resonances for hybridizing the bands with a single harmonic waveform.
An estimate of the gap energy is derived using a Landau-Zener formula on the measured transferred fraction \cite{Suppl}.
The data is presented in Fig.~\ref{fig2}(b-d).
For comparison, we show the evaluation from numerical Floquet-Bloch simulations as solid lines in the figure plots. 
To cover a larger range of energies, we use different Bloch oscillation frequencies $\nu_\mathrm{BO}$.
In Fig.~\ref{fig2}(b) we measure the linear dependence of the gap on the driving strength in a direct resonance situation $\hbar\omega=\Delta(q_c)$ (one-photon transition).
If the gap becomes much larger than the Bloch oscillation frequency $h \nu_\mathrm{BO}$ no atoms are transferred, independent of the gap size, and a saturation of the data is observed.
Figure~\ref{fig2}(c) shows a two-photon resonance $2\hbar\omega=\Delta(q_c)$ with the approximate parabolic opening of the gap versus driving strength $K_\omega$.
A three photon resonance $3\hbar\omega = \Delta(q_c)$ transition is probed and the extracted gap size is plotted in Fig.~\ref{fig2}(d).

Since the coupling strength decreases with the order of the process, we measure overall reduced gap values for equal driving strengths for three-photon processes compared to two- and one-photon processes.
For intermediate driving strengths the data agrees with the theoretical prediction.
Strong driving leads to a fragmentation of the Floquet-Bloch spectrum and the single gap description breaks down.
For deeper lattices the $sp$-gaps scale differently depending on the order of the process.
In general, interband coupling to neighboring sites and hopping beyond nearest neighbors become negligible processes and the static band gap increases leading to larger driving frequencies to match the resonance condition.
While this means that for odd photon number processes the strongest term always scales proportional to $\sim\hbar\omega$, a slight decrease of the interband coupling $\eta_{sp}^{(0)}$ leads overall to larger gaps for the single photon resonance and lower gaps for the three photon resonance at deeper lattices.
In the case of even photon number processes, the strongest process is independent of the driving frequency and scales with the tunneling matrix element.
Therefore, the two-photon gap decreases for deeper lattices.  
However, for all processes a decrease of the coupling can be compensated by enhancing the driving strength because undesired higher band couplings also become less at deeper lattices.

\begin{figure}
    \includegraphics{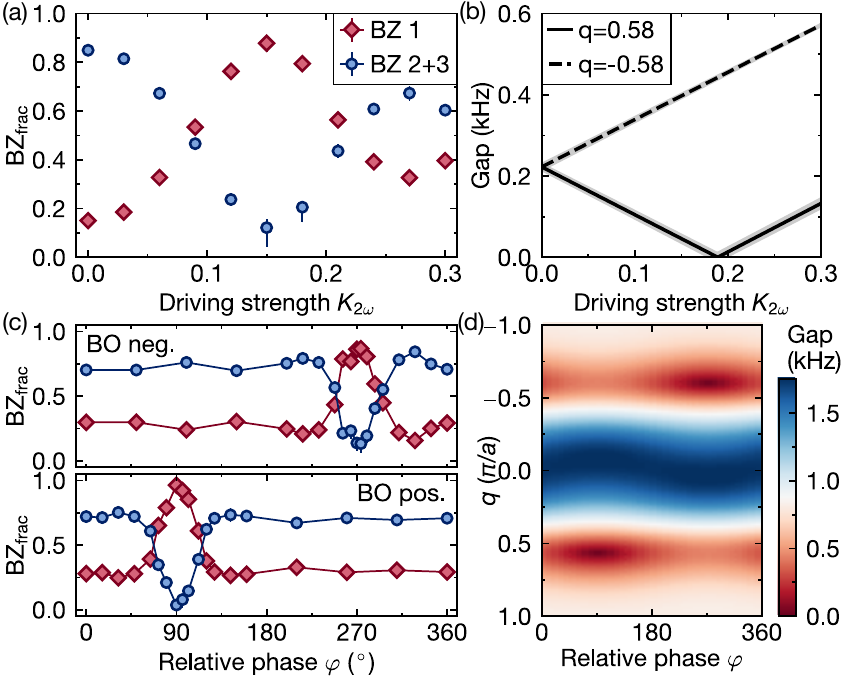}
    \caption{Closing individual gaps. Panel (a) shows the band populations measured after a Bloch oscillation of $\nu_\mathrm{BO}=15\ \mathrm{Hz}$ through half the Brillouin zone for a two-frequency modulated ($\omega/2\pi = 3550\ \mathrm{Hz},2\omega/2\pi = 7100\ \mathrm{Hz}$), one-dimensional lattice of depth $V_X=6.0\ \mathrm{E_R}$. The data is taken at different driving strengths of the higher harmonic $K_{2\omega}$ at fixed $K_\omega=1.0(1)$ and $\varphi=90.0(5)^\circ$. In (b) we plot the gap calculated in an effective Hamiltonian picture at the critical quasimomenta $q_c = \pm 0.587\ \pi/a$. The gap closes and opens linearly with $K_{2\omega}$ for fixed $K_\omega=1.0, \varphi=90^\circ$. Positive and negative quasimomenta can be individually probed by changing the direction of the Bloch oscillation denoted as 'BO pos.' and 'BO neg.' in panel (c). For the driving parameters $K_\omega=1.0(1), K_{2\omega}=0.155(2)$, the band population is measured at various relative phases $\varphi$ of the drive. The calculated size of the gap is shown versus quasimomentum and $\varphi$ in panel (d) for $K_\omega=1.0, K_{2\omega}=0.18$. The y-error bars on the experimental data reflect a statistical standard error of 4 measurements. Statistical and systematic error in the shaking strength and relative phase amount to $1 \%$ and $1^\circ$, respectively. The shaded area for the theoretical curves represents uncertainties in the shaking strength $K_\omega$ and lattice depth. \label{fig3}}
\end{figure}
So far, we have demonstrated the control on the $sp$-band coupling in quantitative agreement using a single frequency driving protocol.
We add a higher harmonic to the drive in order to control individual gaps in the effective Floquet bands.
In the case of a two-photon resonance (Fig.~\ref{fig2}(c)), the gap value of the single-frequency driven lattice reaches about a tenth of the recoil energy using a driving strength of $K_\omega=1.0$ at $\omega/2\pi=3550\ \mathrm{Hz}$.
If we choose the higher harmonic of the drive at exactly twice the frequency ($2\omega/2\pi = 7100\ \mathrm{Hz}$), we can resonantly address the same transition.
The gap size can now also be tuned through two additional parameters, the driving strength $K_{2\omega}$ and the relative phase $\varphi$ between the two harmonics.
As shown in Fig.~\ref{fig1}(c), we restore time-glide symmetry of the driving potential at a relative phase of $\varphi=90^\circ$ and are able to close the gap.
To detect the gap closing we use the same method as for measuring the gap size.
In Fig.~\ref{fig3}(a) we plot the band populations after crossing the transition point.
If the population stays in the initial $s$-band, the gap size is below the resolution limit given by the Bloch oscillation frequency.
We measure the gap closing to occur at $K_{2\omega}=0.155(2)$ which is slightly lower than the theoretical value of $0.18(2)$ shown in Fig.~\ref{fig3}(b).
The perturbative effective Hamiltonian is used for the theoretically obtained data.
The two plotted lines represent the theoretically predicted gap at the two critical quasimomenta ($q_c=\pm 0.585\ \pi/a$) versus driving amplitude of the second harmonic.
For a phase of $\varphi=90^\circ$, the gap closes linearly at positive quasimomentum $q_c$, while a linear opening is induced at the negative quasimomentum $q_c$.

We can individually probe both gaps by reversing the magnetic field gradient and inducing a Bloch oscillation in the opposite direction.
Choosing the previously measured critical strength of $K_{2\omega}=0.155$, we scan the relative phase and measure the band populations after moving through the Brillouin zone, as shown in Fig.~\ref{fig3}(c).
Bloch oscillations to the left probe the negative branch of quasimomenta ($q \in [-1,0]$) and the gap closing is detected at $\varphi=271(2)^\circ$.
Likewise, Bloch oscillations to the right probe the positive branch of quasimomenta ($q \in [0,1]$) and the gap closing is detected at $\varphi=92(1)^\circ$.
We calculate the expected gap with quasi-degenerate perturbation theory and plot the results versus quasimomentum and relative phase in Fig.~\ref{fig3}(d).
For this calculation we use the critical driving strengths $K_\omega=1.0, K_{2\omega}=0.18$.
At these strengths the gaps close at the corresponding quasimomenta but stay finite elsewhere.
The finite Bloch oscillation frequency gives a lower bound on the minimal gap.
However, the frequency is chosen such that on typical experimental timescales ($\simeq 100\ \mathrm{ms}$) the gap is effectively closed.

\begin{figure}
    \includegraphics{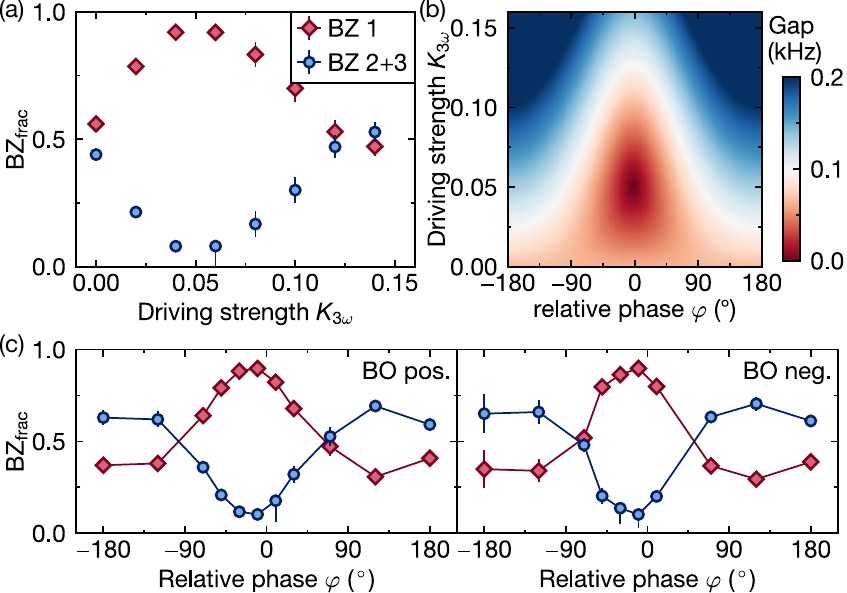}
    \caption{Closing of both gaps. In (a) we probe the gap opened by the two-frequency driving of a one-dimensional lattice using Bloch oscillations with frequency $\nu_\mathrm{BO}=15\ \mathrm{Hz}$. The band population is measured against the driving strength $K_{3\omega}$ for fixed modulation parameters $K_\omega=1.50(2), \varphi=0(1)^\circ, \omega/2\pi=2440\ \mathrm{Hz}, 3\omega/2\pi=7320\ \mathrm{Hz}$. A low fraction of transferred atoms indicates a closing of the gap. A theoretical estimation of the gap is plotted in (b) at fixed $K_\omega=1.5$ for the same driving frequencies as above. We use quasidegenerate perturbation theory to derive the effective model. Panel (c) shows the experimental measurements of the individual gap closings at positive and negative quasimomenta. The extremal points in the band population occur at a relative phase of $\varphi=-17(4)^\circ$. Error bars on the extracted populations combine a statistical standard error of four measurements with systematic errors from the fitting protocol. Statistical and systematic error in the shaking strength and relative phase add up to $1 \%$ and $1^\circ$, respectively. \label{fig4}}
\end{figure}
Changing the ratio of driving frequencies from $1/2$ to $1/3$ enables us to simultaneously control both gaps.
To do so, we drive resonantly a three-photon transition in combination with a second frequency that directly adresses the same transition, therefore, synthesizing the fundamental and third harmonic in the modulation waveform.
The fundamental frequency is fixed to $\omega/2\pi=2440\ \mathrm{Hz}$ with  strength $K_\omega=1.5$ which induces a gap of roughly $75$~Hz at the transition point.
In Fig.~\ref{fig4}(a) the band population is measured as a function of the driving strength $K_{3\omega}$ of the third harmonic $3\omega/2\pi=7320\ \mathrm{Hz}$ with relative phase $\varphi=0^\circ$.
Since the gap is smaller compared to the two-photon resonance, the strength of the additional driving needed to close the gap is also weaker.
We estimate a critical driving strength of $K_{3\omega}=0.06(1)$ to close the gap.
The gap size calculated by the perturbative model is shown versus the two driving parameters introduced by the third harmonic ($K_{3\omega},\varphi$) in Fig.~\ref{fig4}(b), in which the minimum gap over the full Brillouin zone is plotted.
For both negative and positive quasimomenta, the gap closes at the same relative phase $\varphi=0^\circ$ for $K_\omega=1.5$ and $K_{3\omega}=0.05$.
In contrast to the two-photon case, the theoretically estimated critical strength matches the experimentally measured one.
The band population measurements versus relative phase are shown in Fig.~\ref{fig4}(c), where the left panel corresponds to the positive and the right panel to the negative quasimomenta.
We estimate the measured minimal transfer at a phase of $\varphi=-17(4)^\circ$ which deviates from the expected minimum at $\varphi=0^\circ$.
However, this can be attributed to a systematic shift in the calibration of the relative phase $\varphi$ at small driving strengths of $K_{3\omega}$.

In this experiment, we have demonstrated full control over individual gaps in $sp$-hybridized Floquet bands.
We developed an effective model that quantitatively agrees with experimental data for both single and multi-frequency driving.
This constitutes a flexible platform to explore and test theories and predictions in the simple setting of one-dimensional lattices.
The scheme extends the possibilities of studying topology in various ladder models \cite{Li2013}, such as the Creutz-ladder model \cite{Kang2018,Kang2020a} or the inversion symmetric Shockley model \cite{Fuchs2021}, where in a tight-binding picture the $s$- and $p$-band correspond to the legs of the ladder.
Although the two-band model does not include the non-negligible couplings between the $p$-dominated effective Floquet band and the higher $d$-band, it accurately models the $s$-dominated effective Floquet band in which we are interested.
In particular it has been proposed how a similar two-frequency scheme can be used to adiabatically prepare a topological insulator from an initially  trivial band insulator of fermions in a simple one-dimensional lattice \cite{Sun2017,Kang2020}.
The creation of an asymmetric band via the closing of only a single gap can be used to create an analogue of an one-dimensional helical edge state \cite{Budich2017}.
Additionally, we are able to use fermionic potassium atoms with the same experimental setup which allows us to introduce tunable on-site interaction via a Feshbach resonance \cite{Kohl2005,Chin2010}.
In the case of hybridized bands the interaction leads to a coupling between the two effective bands and further decoupling from third and higher band resonances is needed which could be achieved by dimerization of the lattice.
The closing of both gaps at the same time and therefore suppressing multiphoton resonances can be used to prevent heating in phase modulated lattices even for strongly interacting situations 
\cite{Viebahn2021}.

\begin{acknowledgments}
We thank Alexander~Frank for his contributions and support on the setup of electronic devices, and Andr\'e~Eckardt, Martin~Eckstein, Yuta~Murakami, Manish~Sajnani and  Philipp~Werner for insightful discussions.
This work was partly funded by the SNF (project no.~182650), NCCR-QSIT, QUIC (Swiss State Secretary for Education, Research and Innovation contract no.~15.0019), and ERC advanced grant TransQ (project no.~742579).
\end{acknowledgments}

\renewcommand{\theequation}{A\arabic{equation}}
\makeatletter
\renewcommand{\thefigure}{A\@arabic\c@figure} 
\renewcommand{\thetable}{A\@arabic\c@table} 
\setcounter{figure}{0}  

\appendix
\section{Floquet effective Hamiltonians using quasi-degenerate perturbation theory}

An intuitive way to calculate the effective Hamiltonian of a Floquet system is quasi-degenerate perturbation theory on the extended space \cite{Eckardt2015}, which is equivalent to the high-frequency expansion.
This method has been used for modulated two-level models \cite{Hausinger2010} and for estimating heating effects in driven optical lattices \cite{Weinberg2015}. 
Here, we extend this method to two-color driving waveforms and derive analytical expressions for effective Hamiltonians.
The first step is to formulate the problem in the Floquet extended space where quasi-degenerate perturbation theory is applied.
In addition, it is convenient to separate the intra- and inter-band coupling terms.
This can be achieved by a transformation into the comoving frame.
We use the tight-binding approximation with higher order corrections to arrive at an analytical expression for the effective Hamiltonian.

\subsection{Tight-binding formulation}
We start with the single particle Hamiltonian
\begin{equation}
\hat{H}_{\text{sp}}(\tau) = \frac{\hat{p}^2}{2M} + V \left[\hat{x} - x_0 (\tau)\right], 
\end{equation}
where the phase modulation takes the form
\begin{align}
k_L x_0 (t) &= k_L\left[\Delta L_\omega \cos (\omega \tau) + \Delta L_{l\omega} \cos (l\omega\tau+ \varphi)\right], \nonumber\\
& =\ \frac{2 E_{\text{rec}}}{\pi \hbar \omega} \left[K_\omega \cos\left(\omega \tau\right) +  \frac{K_{l\omega} }{l} \cos\left(l\omega\tau+ \varphi\right)\right].
\end{align}
The expansion of the piezo-electric actuator $\Delta L_{l\omega}$ is rewritten in terms of a dimensionless parameter $K_{l\omega}$ representing the strength of the modulation.
The second harmonic component is a multiple of the basic driving frequency with multiple $l \in \mathbb{N}$.
After transforming to the comoving frame
\begin{equation}
\hat{H}^{\prime}_{\text{sp}}(\tau) = \frac{\hat{p}^2}{2m} + V \left(\hat{x}\right) + M\ddot{x}_0(\tau)\hat{x},
\end{equation}
the Hamiltonian is no longer translational invariant. 
However, we still can exploit the Bloch theorem and find Bloch states $\psi_{q\left(\tau\right),n}$ with band index $n$ but for a time-dependent quasimomentum
\begin{equation} 
q\left(\tau\right)=q - \frac{M}{\hbar} \dot{x}_0 \left(\tau\right).
\end{equation}
In second quantization, the Hamiltonian can be written as
\begin{align}\label{eqn:timedepH}
\hat{H}(\tau) =& \sum_{q\left(t\right),n} \varepsilon_n \left[q\left(\tau\right)\right] \hat{c}_{q\left(\tau\right),n}^\dagger \hat{c}_{q\left(\tau\right),n}\nonumber\\
&+ M\ddot{x}_0(\tau)\sum_{q\left(\tau\right),n,n^\prime} \eta_{n n^\prime}\left[q\left(\tau\right)\right] \hat{c}_{q\left(\tau\right),n}^\dagger \hat{c}_{q\left(\tau\right),n^\prime},
\end{align}
with $\hat{c}_{q\left(\tau\right),n}$ the annihilation operator for a Bloch state $\psi_{q\left(\tau\right),n}$, $\varepsilon_n \left[q\left(\tau\right)\right]$ the dispersion matrix element and $\eta_{n n^\prime}\left[q\left(\tau\right)\right]$ the dipole matrix element.\\
We use a tight-binding approximation to find analytical expressions for the matrix element of the dispersion
\begin{align}\label{eqn:timedepEps}
\varepsilon_n \left[q-\frac{M \dot{x}_0}{\hbar}\right]\simeq\ & E_n +\sum_{p=1}^P 2t^{(p)}_n \cos\left[pa\left(q-\frac{M \dot{x}_0}{\hbar}\right)\right], 
\end{align}
where $E_n$ is the bandcenter energy and $t_n^{(p)}$ are the static tunneling matrix elements starting of order $p$, where $p=1$ corresponds to nearest neighbor hopping. 
The maximum order of the tight-binding approximation is defined by $P$.
The numerical values of $E_n$ and $t_n^{(p)}$ are obtained by Fourier decomposition of the time-independent dispersion, see Table~\ref{tab:TBparams}.
In the same manner the dipole matrix element can be expanded in a Fourier series in quasimomentum
\begin{align}\label{eqn:timedepEta}
\eta_{n n^\prime} \left[q-\frac{M \dot{x}_0}{\hbar}\right]\simeq\ & \eta_{n n^\prime}^{(0)} +\sum_{p=1}^P \eta_{n n^\prime}^{(p)} \left( e^{ipaq}e^{-ipa M \dot{x}_0/\hbar}\right. \nonumber \\
& - (-1)^{\mathcal{P}_n+\mathcal{P}_{n^\prime}} \left. e^{-ipaq}e^{ipa M \dot{x}_0/\hbar}\right),
\end{align}
where the expansion is either an even or odd function in quasimomentum depending on the parity of the bands
\begin{equation}
\mathcal{P}_n = \begin{cases}
+1, & \text{if }n=s,d,\dotsc,\\
-1, & \text{if }n=p,f,\dotsc.
\end{cases}
\label{eqn:parity}
\end{equation}
The Fourier coefficients of the static dipole matrix elements $\eta_{n,n^\prime}^{\left(p\right)}$ for couplings between band $n$ and $n^\prime$ are calculated in the Bloch basis using its relation to the momentum operator \cite{Gu2013},
\begin{equation}
\langle \psi_{n q} | \hat{x} | \psi_{n^\prime q^\prime} \rangle = -\frac{i\hbar}{M \left(\varepsilon_{n q}-\varepsilon_{n^\prime,q^\prime}\right)} \langle \psi_{nq} | \hat{p} | \psi_{n^\prime q^\prime} \rangle.
\end{equation}
The results are presented in Table~\ref{tab:TBparams}.
The first order term ($p=1$) corresponds to an interband coupling between neighboring sites.
\begin{table*}
\caption{\label{tab:TBparams}Tight-binding parameters for a $V_X=6\ \text{E}_\text{R}$ lattice, where we use the recoil energy $\text{E}_\text{R}=2026$~Hz for $^{87}$Rb in a retro-reflected 1064~nm optical lattice. The values are computed including 30~bands and sampling the Brillouin zone with 1001~quasimomentum values.}
\begin{ruledtabular}
\begin{tabular}{crrrcrrr}
&\multicolumn{3}{c}{band $n$}& & \multicolumn{3}{c}{coupling $n n^\prime$}\\
&s&p&d& &sp&sd&pd \\ \hline
$E_n/h$ (Hz)&$-1700$&$5754$&$13393$&$\eta_{n n^\prime}^{(0)}$&$0.16407$&$0.00000$&$-0.24241$ \\
$t_n^{(1)}/h$ (Hz)&$-103$&$779$&$-1888$&$\eta_{n n^\prime}^{(1)}$&$-0.03141$&$-0.01504$&$-0.16946$\\
$t_n^{(2)}/h$ (Hz)&$4$&$128$&$169$&$\eta_{n n^\prime}^{(2)}$&$0.00330$&$-0.00342$&$-0.11246$ \\
$t_n^{(3)}/h$ (Hz)&$-0$&$45$&$-177$&$\eta_{n n^\prime}^{(3)}$&$0.00001$&$-0.00343$&$-0.07592$ \\
$t_n^{(4)}/h$ (Hz)&$0$&$18$&$56$&$\eta_{n n^\prime}^{(4)}$&$0.00034$&$-0.00048$&$-0.05102$\\
$t_n^{(5)}/h$ (Hz)&$0$&$9$&$-56$&$\eta_{n n^\prime}^{(5)}$&$0.00020$&$-0.00131$&$-0.03414$\\
\end{tabular}
\end{ruledtabular}
\end{table*}

\subsection{Extended space}
The time periodicity of the problem is exploited by combining the Hilbert space $\mathcal{H}$ of the Bloch functions with the space of square integrable, T-periodic functions $\mathcal{L}_T$ to the extended Hilbert space $\mathcal{F}=\mathcal{H}\otimes \mathcal{L}_T$ \cite{Eckardt2015}.
An orthonormal basis set in this space is acquired by extending the Bloch basis to
\begin{equation}
|nqm\rangle = |nq\rangle e^{im\omega \tau}.
\end{equation}
We call the additional state index $m$ of the extended basis 'photon number'. 
The time-dependent Schr\"odinger equation can be written in the eigenvalue problem for the quasienergy operator
\begin{equation}
\hat{Q}|nqm\rangle = \tilde{\varepsilon}_{nm} |nqm\rangle,
\end{equation}
where the matrix elements of this operator are given by the Fourier coefficients
\begin{equation}
\hat{H}_m = \frac{1}{T} \int_0^T \hat{H}(\tau) e^{-im\omega \tau} d\tau
\end{equation}
of the time-dependent Hamiltonian of Eq.~\ref{eqn:timedepH},
\begin{align}
\langle n^\prime q^\prime m^\prime | \hat{Q} | n q m \rangle = & \langle n^\prime q^\prime | \hat{H}_{m-m^\prime}| n q \rangle \nonumber\\
&+\delta_{mm^\prime}\delta_{qq^\prime} \left(m-m^\prime\right) \hbar \omega.
\end{align}
Since the Hamiltonian is already diagonal in $q$, we drop the $q$-dependence in the following notation for clarity.
The intraband ($n=n^\prime$) contributions can be calculated using Eq.~\ref{eqn:timedepEps},
\begin{align}
&\varepsilon_{n,m-m^\prime} = \langle n | \hat{H}_{m-m^\prime} |n\rangle \nonumber\\
& = E_n\delta_{0,m-m^\prime} +\sum_{p=1}^{P} t_n^{(p)}\left(g_{m-m^\prime} e^{ipaq}+g^{\ast}_{m^\prime-m} e^{-ipaq}\right), \label{eqn:eps}
\end{align}
with the Fourier coefficients
\begin{equation}
g_{m-m^\prime} = \frac{1}{T}\int_0^T e^{ipaM\dot{x}(\tau)/\hbar}e^{-i(m-m^\prime)\omega \tau} \mathrm{d}\tau.
\label{eqn:gfourier}
\end{equation}
Similarly, the interband ($n \neq n^\prime$) transitions follow from Eq.~\ref{eqn:timedepEta},
\begin{align}
\eta_{n n^\prime,m-m^\prime} =& \langle n^\prime | \hat{H}_{m-m^\prime}|n \rangle \nonumber\\
=& -\hbar\omega\eta_{nn^\prime}^{(0)} f_{m-m^\prime} -\hbar\omega\sum_{p=1}^{P} \eta_{nn^\prime}^{(p)}\frac{m-m^\prime}{p}\nonumber\\
&\times\left( g_{m-m^\prime} e^{ipaq}-(-1)^{\mathcal{P}_n+\mathcal{P}_{n^\prime}}g^{\ast}_{m^\prime-m} e^{-ipaq}\right),
\label{eqn:eta}
\end{align}
with the Fourier coefficients
\begin{align}
f_{m-m^\prime}=&\frac{K_\omega}{2}\left(\delta_{1,m-m^\prime}+\delta_{-1,m-m^\prime}\right)\nonumber\\
&+\frac{lK_{l\omega}}{2}\left(\delta_{l,m-m^\prime}e^{i\varphi}+\delta_{-l,m-m^\prime}e^{-i\varphi}\right).
\end{align}
In order to find the matrix elements \ref{eqn:eps} and \ref{eqn:eta} we can use the Jacobi-Anger expansion \cite{Abramowitz1964} for Eq.~\ref{eqn:gfourier}
\begin{align}
e^{-ipaM \dot{x}_0/\hbar} =\sum_{r,r^\prime=-\infty}^{\infty} \mathcal{J}_r \left(pK_\omega\right) \mathcal{J}_{r^\prime} \left(pK_{l\omega}\right)\nonumber\\
\times  e^{i\omega\tau\left(r+lr^\prime\right)}e^{i r^\prime \phi},
\label{eqn:JacobiAnger}
\end{align}
where $\mathcal{J}_r$ are the Bessel functions of the first kind of order~$r$.
We approximate this sum by neglecting all terms which are lower than $10^{-3}$ in magnitude using the driving parameters in the paper.
The results of the quasienergy matrix elements are presented in Table~\ref{tab:JacobiAnger}. \\ 
\begin{table*}
\caption{\label{tab:JacobiAnger}Matrix elements of the quasienergy operator $\hat{Q}$ from Eq.~\ref{eqn:eps} and \ref{eqn:eta} in the Floquet extended space $\mathcal{F}$ with photon number difference $\Delta m = m-m^\prime$. A tight-binding approximation of order $P$ is used to describe the dispersion of band $n$ in the one-dimensional lattice using the band center energies $E_n$, tunneling elements $t_n^{(p)}$ and dipole matrix elements $\eta_{n,n^\prime}^{(p)}$. The sum in Eq.~\ref{eqn:JacobiAnger} is approximated to only include terms larger than $10^{-3}$ with respect to the largest one for the driving strengths used.}
\begin{ruledtabular}
\begin{tabular}{lll}
 \multicolumn{3}{c}{Single frequency}\\
 $\Delta m$& intraband coupling $\varepsilon_{n,m-m^\prime}$& interband coupling $\eta_{n n^\prime,m-m^\prime}$\\ \hline
 0&$E_n +\sum_{p=1}^P 2 t_n^{(p)} \cos\left(paq\right)\mathcal{J}_0\left(pK_\omega\right)$&0\\
 1&$\sum_{p=1}^P 2 t_n^{(p)} i\sin\left(paq\right)\mathcal{J}_1\left(pK_\omega\right)$&$-\hbar\omega \left\{\frac{1}{2}K_\omega\eta_{nn^\prime}^{(0)} +\sum_{p=1}^P \frac{1}{p}\eta_{nn^\prime}^{(p)}\mathcal{J}_1(pK_\omega)\left[e^{ipaq}+(-1)^{\mathcal{P}_n+\mathcal{P}_{n^\prime}}e^{-ipaq}\right]\right\}$\\
 2&$\sum_{p=1}^P 2 t_n^{(p)} \cos\left(paq\right)\mathcal{J}_2\left(pK_\omega\right)$&$-\hbar\omega \sum_{p=1}^P \frac{2}{p}\eta_{nn^\prime}^{(p)}\mathcal{J}_2(pK_\omega)\left[e^{ipaq}-(-1)^{\mathcal{P}_n+\mathcal{P}_{n^\prime}}e^{-ipaq}\right]$\\
 3&$\sum_{p=1}^P 2 t_n^{(p)} i\sin\left(paq\right)\mathcal{J}_3\left(pK_\omega\right)$&$-\hbar\omega \sum_{p=1}^P \frac{3}{p}\eta_{nn^\prime}^{(p)}\mathcal{J}_3(pK_\omega)\left[e^{ipaq}+(-1)^{\mathcal{P}_n+\mathcal{P}_{n^\prime}}e^{-ipaq}\right]$\\
 4& $\sum_{p=1}^P 2 t_n^{(p)} \cos\left(paq\right)\mathcal{J}_4\left(pK_\omega\right)$&$-\hbar\omega \sum_{p=1}^P \frac{4}{p}\eta_{nn^\prime}^{(p)}\mathcal{J}_4(pK_\omega)\left[e^{ipaq}-(-1)^{\mathcal{P}_n+\mathcal{P}_{n^\prime}}e^{-ipaq}\right]$\\ \hline
 \hline
 \multicolumn{3}{c}{Two frequency - $l=2$}\\
 $\Delta m$& \multicolumn{2}{l}{intraband coupling $\varepsilon_{n,m-m^\prime}$}\\ \hline
 0&\multicolumn{2}{l}{$E_n + \sum_{p=1}^P 2 t_n^{(p)}\left\{ \cos\left(paq\right)\mathcal{J}_0\left(pK_\omega\right)\mathcal{J}_0\left(pK_{2\omega}\right)- \sin\left(paq\right)2\mathcal{J}_2\left(pK_\omega\right)\mathcal{J}_1\left(pK_{2\omega}\right)\sin(\varphi)\right\}$}\\
 1&\multicolumn{2}{l}{$\sum_{p=1}^P 2 t_n^{(p)}\left\{i\sin\left(paq\right)\mathcal{J}_1\left(pK_\omega\right)\mathcal{J}_0\left(pK_{2\omega}\right)-\cos\left(paq\right)\left[\mathcal{J}_1\left(pK_{\omega}\right)\mathcal{J}_1\left(pK_{2\omega}\right)e^{i\varphi}+\mathcal{J}_3\left(pK_{\omega}\right)\mathcal{J}_1\left(pK_{2\omega}\right)e^{-i\varphi}\right]\right\}$}\\
 2&\multicolumn{2}{l}{$\sum_{p=1}^P 2 t_n^{(p)}\left\{\cos\left(paq\right)\left[\mathcal{J}_2\left(pK_{\omega}\right)\mathcal{J}_0\left(pK_{2\omega}\right)+\mathcal{J}_2\left(pK_{\omega}\right)\mathcal{J}_2\left(pK_{2\omega}\right)e^{i2\varphi}\right]\right.$}\\
  &\multicolumn{2}{l}{$\left. +i\sin\left(paq\right)\left[\mathcal{J}_0\left(pK_{\omega}\right)\mathcal{J}_1\left(pK_{2\omega}\right)e^{i\varphi}-\mathcal{J}_4\left(pK_{\omega}\right)\mathcal{J}_1\left(pK_{2\omega}\right)e^{-i\varphi}\right]\right\}$}\\
 3&\multicolumn{2}{l}{$\sum_{p=1}^P 2 t_n^{(p)}\left\{i\sin\left(paq\right)\left[\mathcal{J}_3\left(pK_\omega\right)\mathcal{J}_0\left(pK_{2\omega}\right)-\mathcal{J}_1\left(pK_\omega\right)\mathcal{J}_2(pK_{2\omega})e^{i2\varphi}\right]+\cos\left(paq\right)\mathcal{J}_1\left(pK_{\omega}\right)\mathcal{J}_1\left(pK_{2\omega}\right)e^{i\varphi}\right\}$}\\
 4&\multicolumn{2}{l}{$\sum_{p=1}^P 2 t_n^{(p)}\left\{\cos\left(paq\right)\left[\mathcal{J}_4\left(pK_{\omega}\right)\mathcal{J}_0\left(pK_{2\omega}\right)+\mathcal{J}_0 \left(pK_{\omega}\right)\mathcal{J}_2 \left(pK_{2\omega}\right)e^{i2\varphi}\right]+i\sin\left(paq\right)\mathcal{J}_2\left(pK_\omega\right)\mathcal{J}_1\left(pK_{2\omega}\right)e^{i\varphi}\right\}$ }\\
\hline
 $\Delta m$& \multicolumn{2}{l}{interband coupling $\eta_{n n^\prime,m-m^\prime}$}\\ \hline
 0&\multicolumn{2}{l}{0}\\
 1&\multicolumn{2}{l}{$-\hbar\omega\frac{1}{2} K_\omega\eta_{n n^\prime}^{(0)}-\hbar\omega\sum_{p=1}^P \frac{1}{p}\eta_{n n^\prime}^{(p)}\left\{\left[e^{ipaq}+(-1)^{\mathcal{P}_n+\mathcal{P}_{n^\prime}}e^{-ipaq}\right]\mathcal{J}_1\left(pK_\omega\right)\mathcal{J}_0\left(pK_{2\omega}\right)\right.$}\\
 &\multicolumn{2}{l}{$\left.-\left[e^{ipaq}-(-1)^{\mathcal{P}_n+\mathcal{P}_{n^\prime}}e^{-ipaq}\right]\left[\mathcal{J}_1\left(pK_{\omega}\right)\mathcal{J}_1\left(pK_{2\omega}\right)e^{i\varphi}+\mathcal{J}_3\left(pK_{\omega}\right)\mathcal{J}_1\left(pK_{2\omega}\right)e^{-i\varphi}\right]\right\}$}\\
 2&\multicolumn{2}{l}{$-\hbar\omega K_{2\omega} \eta_{n n^\prime}^{(0)}e^{i\varphi}-\hbar\omega \sum_{p=1}^P\frac{2}{p}\eta_{n n^\prime}^{(p)} \left\{ \left[e^{ipaq}-(-1)^{\mathcal{P}_n+\mathcal{P}_{n^\prime}}e^{-ipaq}\right]\left[\mathcal{J}_2\left(pK_{\omega}\right)\mathcal{J}_0\left(pK_{2\omega}\right)+\mathcal{J}_2\left(pK_{\omega}\right)\mathcal{J}_2\left(pK_{2\omega}\right)e^{i2\varphi}\right]\right.$}\\
 &\multicolumn{2}{l}{$\left.+\left[e^{ipaq}+(-1)^{\mathcal{P}_n+\mathcal{P}_{n^\prime}}e^{-ipaq}\right]\left[\mathcal{J}_0\left(pK_{\omega}\right)\mathcal{J}_1\left(pK_{2\omega}\right)e^{i\varphi}-\mathcal{J}_4\left(pK_{\omega}\right)\mathcal{J}_1\left(pK_{2\omega}\right)e^{-i\varphi}\right]\right\}$}\\
 3&\multicolumn{2}{l}{$-\hbar\omega \sum_{p=1}^P\frac{3}{p}\eta_{n n^\prime}^{(p)}\left\{\left[e^{ipaq}+(-1)^{\mathcal{P}_n+\mathcal{P}_{n^\prime}}e^{-ipaq}\right] \left[\mathcal{J}_3\left(pK_\omega\right)\mathcal{J}_0\left(pK_{2\omega}\right)-\mathcal{J}_1\left(pK_\omega\right)\mathcal{J}_2(pK_{2\omega})e^{i2\varphi}\right]\right.$}\\
 &\multicolumn{2}{l}{$\left.+\left[e^{ipaq}-(-1)^{\mathcal{P}_n+\mathcal{P}_{n^\prime}}e^{-ipaq}\right]\mathcal{J}_1\left(pK_{\omega}\right)\mathcal{J}_1\left(pK_{2\omega}\right)e^{i\varphi}\right\}$}\\
 4&\multicolumn{2}{l}{$-\hbar\omega \sum_{p=1}^P \frac{4}{p}\eta_{n n^\prime}^{(p)} \left\{\left[e^{ipaq}-(-1)^{\mathcal{P}_n+\mathcal{P}_{n^\prime}}e^{-ipaq}\right]\left[\mathcal{J}_4\left(pK_{\omega}\right)\mathcal{J}_0\left(pK_{2\omega}\right)+\mathcal{J}_0\left(pK_{\omega}\right)\mathcal{J}_2\left(pK_{2\omega}\right)e^{i2\varphi}\right]\right.$ }\\
 &\multicolumn{2}{l}{$\left.+\left[e^{ipaq}+(-1)^{\mathcal{P}_n+\mathcal{P}_{n^\prime}}e^{-ipaq}\right]\mathcal{J}_2\left(pK_\omega\right)\mathcal{J}_1\left(pK_{2\omega}\right)e^{i\varphi}\right\}$}\\ \hline
 \hline
 \multicolumn{3}{c}{Two frequency - $l=3$}\\
 $\Delta m$&\multicolumn{2}{l}{ intraband coupling $\varepsilon_{n,m-m^\prime}$}\\ \hline
 0&\multicolumn{2}{l}{$E_n + \sum_{p=1}^P 2 t_n^{(p)} \cos\left(paq\right)\left[\mathcal{J}_0\left(pK_\omega\right)\mathcal{J}_0\left(pK_{3\omega}\right)-2\mathcal{J}_3\left(pK_\omega\right)\mathcal{J}_1\left(pK_{3\omega}\right)\cos(\varphi)\right]$ }\\
 1&\multicolumn{2}{l}{$\sum_{p=1}^P 2 t_n^{(p)}i\sin\left(paq\right)\left[\mathcal{J}_1\left(pK_\omega\right)\mathcal{J}_0\left(pK_{3\omega}\right)+\mathcal{J}_2\left(pK_\omega\right)\mathcal{J}_1\left(pK_{3\omega}\right)e^{i\varphi}-\mathcal{J}_4\left(pK_\omega\right)\mathcal{J}_1\left(pK_{3\omega}\right)e^{-i\varphi}\right]$}\\
 2&\multicolumn{2}{l}{$\sum_{p=1}^P 2 t_n^{(p)} \cos\left(paq\right)\left[\mathcal{J}_2\left(pK_\omega\right)\mathcal{J}_0\left(pK_{3\omega}\right)-\mathcal{J}_1\left(pK_\omega\right)\mathcal{J}_1\left(pK_{3\omega}\right)e^{i\varphi}\right]$}\\
 3&\multicolumn{2}{l}{$\sum_{p=1}^P 2 t_n^{(p)}i \sin\left(paq\right)\left[\mathcal{J}_3\left(pK_\omega\right)\mathcal{J}_0\left(pK_{3\omega}\right)+\mathcal{J}_0\left(pK_\omega\right)\mathcal{J}_1\left(pK_{3\omega}\right)e^{i\varphi}\right]$}\\
 4&\multicolumn{2}{l}{$\sum_{p=1}^P 2 t_n^{(p)}\cos\left(paq\right)\left[\mathcal{J}_4\left(pK_\omega\right)\mathcal{J}_0\left(pK_{3\omega}\right)+\mathcal{J}_1\left(pK_\omega\right)\mathcal{J}_1\left(pK_{3\omega}\right)e^{i\varphi}+\mathcal{J}_2\left(pK_\omega\right)\mathcal{J}_2\left(pK_{3\omega}\right)e^{i2\varphi}\right]$ }\\ \hline
 $\Delta m$&\multicolumn{2}{l}{ interband coupling $\eta_{n n^\prime,m-m^\prime}$}\\ \hline
 0&\multicolumn{2}{l}{0}\\
 1&\multicolumn{2}{l}{$-\hbar\omega\frac{1}{2}K_\omega \eta_{n n^\prime}^{(0)} $}\\
 &\multicolumn{2}{l}{$-\hbar\omega \sum_{p=1}^P \frac{1}{p} \eta_{nn^\prime}^{(p)} \left[e^{ipaq}+(-1)^{\mathcal{P}_n+\mathcal{P}_{n^\prime}}e^{-ipaq}\right]\left[\mathcal{J}_1\left(pK_\omega\right)\mathcal{J}_0\left(pK_{3\omega}\right)+\mathcal{J}_2\left(pK_\omega\right)\mathcal{J}_1\left(pK_{3\omega}\right)e^{i\varphi}-\mathcal{J}_4\left(pK_\omega\right)\mathcal{J}_1\left(pK_{3\omega}\right)e^{-i\varphi}\right] $}\\
 2&\multicolumn{2}{l}{$-\hbar\omega\sum_{p=1}^P \frac{2}{p} \eta_{nn^\prime}^{(p)}\left[e^{ipaq}-(-1)^{\mathcal{P}_n+\mathcal{P}_{n^\prime}}e^{-ipaq}\right]\left[\mathcal{J}_2\left(pK_\omega\right)\mathcal{J}_0\left(pK_{3\omega}\right)-\mathcal{J}_1\left(pK_\omega\right)\mathcal{J}_1\left(pK_{3\omega}\right)e^{i\varphi}\right]$}\\
 3&\multicolumn{2}{l}{$-\hbar\omega\frac{3}{2}K_{3\omega} \eta_{n n^\prime}^{(0)}e^{i\varphi} -\hbar\omega \sum_{p=1}^P \frac{3}{p} \eta_{nn^\prime}^{(p)} \left[e^{ipaq}+(-1)^{\mathcal{P}_n+\mathcal{P}_{n^\prime}}e^{-ipaq}\right]\left[\mathcal{J}_3\left(pK_\omega\right)\mathcal{J}_0\left(pK_{3\omega}\right)+\mathcal{J}_0\left(pK_\omega\right)\mathcal{J}_1\left(pK_{3\omega}\right)e^{i\varphi}\right]$}\\
 4&\multicolumn{2}{l}{$-\hbar\omega\sum_{p=1}^P\frac{4}{p}\eta_{nn^\prime}^{(p)} \left[e^{ipaq}-(-1)^{\mathcal{P}_n+\mathcal{P}_{n^\prime}}e^{-ipaq}\right]\left[\mathcal{J}_4\left(pK_\omega\right)\mathcal{J}_0\left(pK_{3\omega}\right)+\mathcal{J}_1\left(pK_\omega\right)\mathcal{J}_1\left(pK_{3\omega}\right)e^{i\varphi}+\mathcal{J}_2\left(pK_\omega\right)\mathcal{J}_2\left(pK_{3\omega}\right)e^{i2\varphi}\right]$ }\\
\end{tabular}
\end{ruledtabular}
\end{table*}

For a given $q$ the block matrix form of $\hat{Q}$ written with the Fourier coefficients of the Hamiltonian takes the form
\begin{equation}
\label{eqn:Qmatrix}
\hat{Q} = \begin{bmatrix}
 \ddots	& 		& \vdots&		&\iddots \\
&  \hat{H}_0-\hbar\omega	& \hat{H}_{1}	& \hat{H}_{2}	 &		\\
\cdots&  \hat{H}_{-1}	& \hat{H}_0	&  \hat{H}_{1}  & \cdots\\
 		&  \hat{H}_{-2}	& \hat{H}_{-1}& \hat{H}_0+\hbar\omega &	\\
\iddots & 		& \vdots&		& \ddots\\
\end{bmatrix},
\end{equation}\\
with the block matrices
\begin{equation}
\label{eqn:QmatrixBlocks}
\hat{H}_{m-m^\prime} = \begin{bmatrix}
 \varepsilon_{s,m-m^\prime}&\eta_{sp,m-m^\prime}& \eta_{sd,m-m^\prime}&\cdots \\
 \eta_{sp,m-m^\prime}^\ast & \varepsilon_{p,m-m^\prime}& \eta_{pd,m-m^\prime}& \\
\eta_{sd,m-m^\prime}^\ast& \eta_{pd,m-m^\prime}^\ast& \varepsilon_{d,m-m^\prime}& \\
\vdots &  & & \ddots	\\
\end{bmatrix},
\end{equation}
where we have labelled the lowest three bands as $n=s,p,d$.
The negative Fourier coefficients ($|m-m^\prime| < 0$) are related by complex conjugation to the positive Fourier coefficients ($|m-m^\prime| > 0$).

\subsection{Quasi-degenerate perturbation theory}
The unperturbed system is given by the time-averaged Hamiltonian $H_0$ and its photon copies $H_0 + m  \hbar\omega$.
The diagonal elements $\varepsilon_{n,0}- m \hbar \omega$ correspond to the static bands and their photon copies which are renormalized by the driving.
Any set of unperturbed states, which are degenerate or almost degenerate, forms a subsystem that typically is energetically separated from the rest of the system (virtual states).
More precisely, this is the case if the driving frequency is a large energy scale compared to the width of the bands of interest.
The unperturbed states are coupled via inter- and intraband transitions and their combination.
However, very high order resonances (photon number difference $m-m^\prime$ of two degenerate states is large) are very weak and can often be neglected for experimentally relevant time scales.

The quasienergy operator is then block diagonalized with respect to the blocks of nearly degenerate states in perturbative fashion.
The resulting effective matrix elements are given by the expansion
\begin{align}
\tilde{\varepsilon}_n =& \tilde{\varepsilon}_n^{(0)} + \tilde{\varepsilon}_n^{(1)} + \tilde{\varepsilon}_n^{(2)} + \dotsb, \\
\tilde{\eta}_{nn^\prime} =& \tilde{\eta}_{nn^\prime}^{(0)} + \tilde{\eta}_{nn^\prime}^{(1)} + \tilde{\eta}_{nn^\prime}^{(2)} + \dotsb,
\end{align}
where the different orders are computed according to quasi-degenerate perturbation theory, see for example \cite{Eckardt2015,Weinberg2015} or Appendix B of \cite{Winkler2003}.\\
\begin{figure}
    \includegraphics{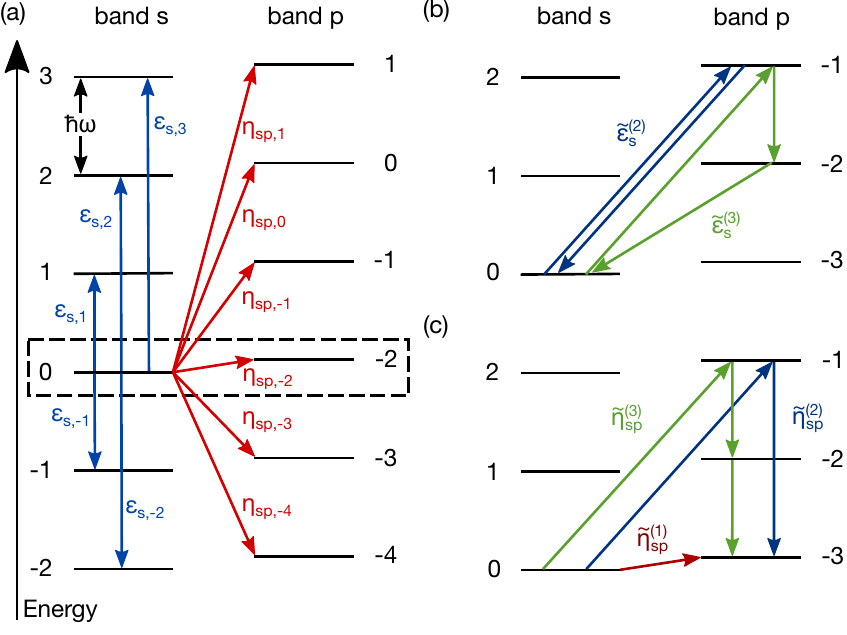}
    \caption{Couplings in the extended Floquet space. Part (a) presents a schematic spectrum of a single unperturbed quasimomentum state in $s$- and $p$-band in the extended space. The dashed box defines a set of quasi-degenerate states for which an effective Hamiltonian can be calculated using perturbation theory. The intraband $\varepsilon_{s,m-m^\prime}$ (blue) and interband $\eta_{sp,m-m^\prime}$ (red) matrix elements of the quasienergy operator connect different photon number ($m,m^\prime$) states indicated by the digit label. Contributions to the effective diagonal $\tilde{\varepsilon}_s$ (b) and off-diagonal $\tilde{\eta}_{sp}$ (c) terms can be visualized as loops and paths involving couplings to virtual states (outside of the dashed box). The different orders of perturbation theory are defined by the amount of virtual states involved and are indicated with different colors.\label{figA1}}
\end{figure}

Figure~\ref{figA1}(a) shows schematically the spectrum of the unperturbed lowest two bands for an individual quasimomentum value in the extended space.
The dashed box indicates a pair of quasi-degenerate states which are mixed via the driving induced coupling.
These two states define the subspace with respect to which we block diagonalize $\hat{Q}$ to get an effective Hamiltonian for these two bands.
All off-diagonal matrix elements of the quasienergy operator $\hat{Q}$ coupling the $|s,0\rangle$ to the other states are depicted as arrows.
Intraband couplings $\varepsilon_{s,m^\prime-m}$ (blue) leave the band index fixed but change the photon number $m$ of the state by $m^\prime-m$.
Typically, small changes in photon number lead to much stronger couplings than larger ones.
On the other hand, interband couplings $\eta_{sp,m^\prime-m}$ (red) change the band index (from $s$ to $p$) with or without shift of the photon number state.
The situation depicted shows a $s$-band state resonant with a two-photon transition to a $p$-band state.
The states outside the box form the virtual states because these states are energetically detuned by the drive energy.

In perturbation theory the effective diagonal and coupling terms can be calculated creating all relevant paths between the quasi-degenerate states.
The paths are built from the coupling elements given by $\hat{Q}$ shown in Fig.~\ref{figA1}(a).
For a contribution to the diagonal effective terms $\tilde{\varepsilon}_n$, all loops are considered, i.e. paths starting and ending at one of the quasi-degenerate states.
An effective coupling between two quasi-degenerate states $\tilde{\eta}_{n n^\prime}$ is composed of all paths that start at the first state and end at the second one.
The order of the perturbation is determined by the amount of virtual states which are included in such a coupling path.
Figure~\ref{figA1}(b) shows an example of first (red), second (blue) and third (green) order contribution to the effective diagonal $s$-band term for a three-photon resonance.
The zeroth order is given by the unperturbed Hamiltonian and by our choice of unperturbed states there is no first order correction to the diagonal terms.
An example of the respective off-diagonal contributions is shown in Fig.~\ref{figA1}(c).
\begin{figure}
    \includegraphics{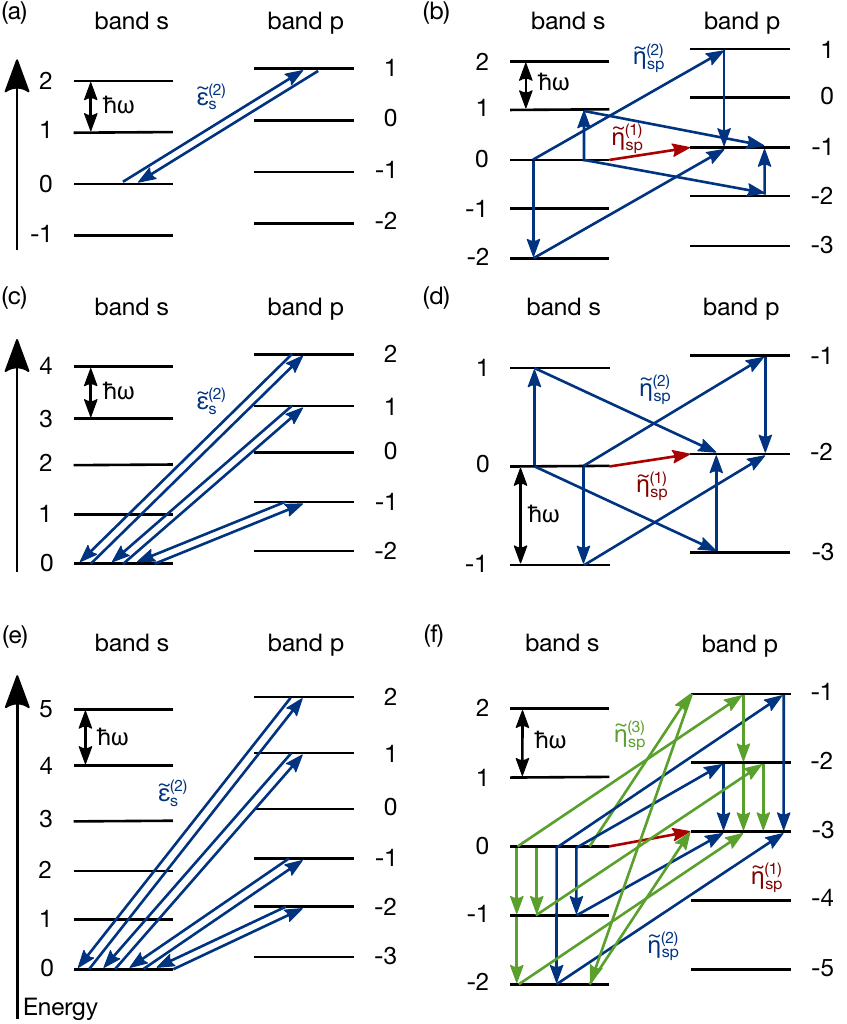}
    \caption{Effective couplings for one- (a,b), two- (c,d) and three-photon (e,f) resonances. The dominant perturbative contributions to the diagonal $\tilde{\varepsilon}_s$ (a,c,e) and off-diagonal $\tilde{\eta}_{sp}$ (b,d,f) terms in the effective two-band Hamiltonian. Only the perturbations on the $s$-band state are shown. The perturbative order of the contributions is indicated by the color. These are the dominant contributions for the parameters given in Table~\ref{tab:TBparams}.\label{figA2}}
\end{figure}

\subsubsection{Effective Hamiltonians}

We benchmark the effective two-band Hamiltonians against the exact spectrum obtained by diagonalization of the one-period evolution matrix \cite{Suppl} for the different driving regimes covered in the main text.
\begin{figure*}
    \includegraphics{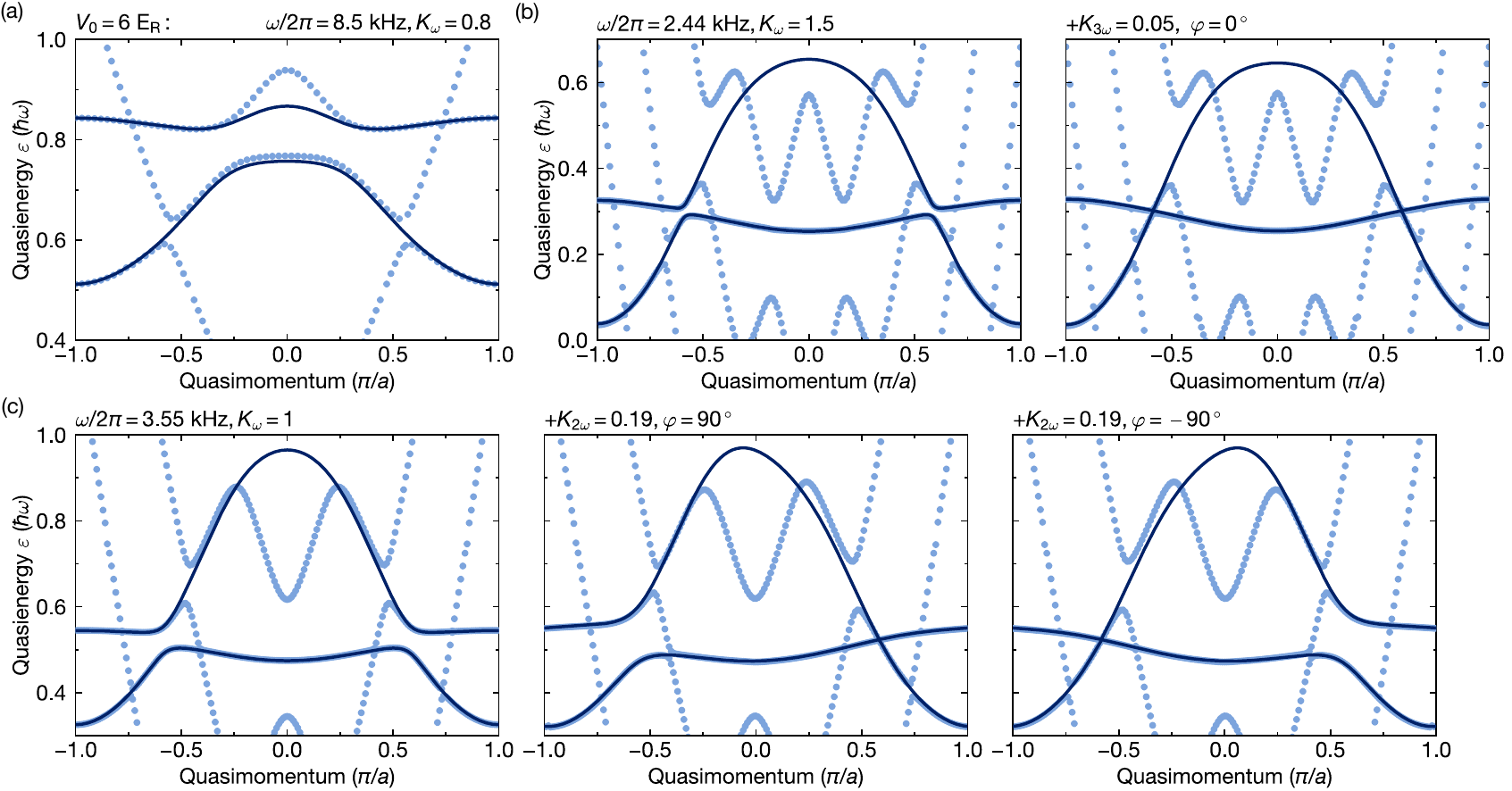}
    \caption{Benchmark effective two-band models. The dispersion of the effective two-band Hamiltonians (solid lines) are compared to numerical exact solution of the Floquet-Bloch band structure (points, three lowest bands). In (a) the two lowest bands tune in resonance with a single photon transition. Part (b) shows the single and  two-frequency driving ($\omega$, $3\omega$) case for a three photon resonance. The first plot in (c) shows a single frequency driving of a two photon resonance, and the other two display the individual gap closings of the two-frequency driving ($\omega$, $2\omega$) case for different relative phases of $\pm 90^\circ$.   \label{figA3}}
\end{figure*}
The basic Hamiltonian is denoted
\begin{equation}
\tilde{H} = 
\begin{pmatrix}
\tilde{\varepsilon}_{s} & \tilde{\eta}_{sp} \\
\tilde{\eta}_{sp}^\ast & \tilde{\varepsilon}_{p}
\end{pmatrix},
\label{eqn:Heff}
\end{equation}
where the tilde indicates that the quantities are results of the perturbation approach.
The results obtained for the quasienergy operator matrix elements (see Table~\ref{tab:JacobiAnger}) together with the relevant paths in perturbation theory (see Fig.~\ref{figA2}) provide the effective matrix elements.

In the case of single frequency driving resonant to the $sp$-band gap, the matrix elements are
\begin{align}
\tilde{\varepsilon}_s = & \varepsilon_{s,0} + \frac{|\eta_{sp,1}|^2}{\delta-2\hbar \omega}, \\
\tilde{\varepsilon}_p = & \varepsilon_{p,0}-\hbar\omega - \frac{|\eta_{sp,1}|^2}{\delta-2\hbar \omega}, \\
\tilde{\eta}_{sp} = & \eta_{sp,1} + \frac{\eta_{sp,2}(\varepsilon_{p,1}^\ast-\varepsilon_{s,1}^\ast)}{2}\left(\frac{1}{\delta+\hbar \omega}+\frac{1}{\hbar \omega}\right)\nonumber\\
& + \frac{\eta_{sp,1}^\ast(\varepsilon_{p,2}-\varepsilon_{s,2})}{2}\left(\frac{1}{\delta-2\hbar\omega}+\frac{1}{2\hbar\omega}\right),	
\end{align}
where $\delta = \varepsilon_{s,0}-\varepsilon_{p,0}+\hbar\omega$ is the detuning from the resonance. Here, we have used the most dominant contributions to the perturbation series shown in Fig.~\ref{figA2}(a)-(b).

At the two-photon resonance, we take into account the effective elements
\begin{align}
\tilde{\varepsilon}_s =\ & \varepsilon_{s,0} + \frac{|\eta_{sp,1}|^2}{\delta+\hbar \omega}+\frac{|\eta_{sp,1}|^2}{\delta-3\hbar\omega}+\frac{|\eta_{sp,2}|^2}{\delta-4\hbar\omega},\\
\tilde{\varepsilon}_p =\ & \varepsilon_{p,0}-2\hbar\omega - \frac{|\eta_{sp,1}|^2}{\delta+\hbar \omega}-\frac{|\eta_{sp,1}|^2}{\delta-3\hbar\omega}-\frac{|\eta_{sp,2}|^2}{\delta-4\hbar\omega},\\
\tilde{\eta}_{sp} =\ & \eta_{sp,2} + \frac{\eta_{sp,1}(\varepsilon_{p,1}-\varepsilon_{s,1})}{2}\left(\frac{1}{\delta-\hbar \omega}-\frac{1}{\hbar \omega}\right)\nonumber\\
& + \frac{\eta_{sp,3}(\varepsilon_{p,1}^\ast-\varepsilon_{s,1}^\ast)}{2}\left(\frac{1}{\delta+\hbar\omega}+\frac{1}{\hbar\omega}\right),
\end{align}
with the two-photon detuning $\delta=\varepsilon_{s,0}-\varepsilon_{p,0}+2\hbar\omega$. The included coupling contributions are shown in Fig.~\ref{figA2}(b)-(c).

The three-photon resonance contribution are depicted in Fig.~\ref{figA2}(e)-(f), and lead to effective elements for the diagonal terms
\begin{align}
\tilde{\varepsilon}_s =\ & \varepsilon_{s,0} + \frac{|\eta_{sp,1}|^2}{\delta-2\hbar \omega}+\frac{|\eta_{sp,1}|^2}{\delta-5\hbar\omega}+\frac{|\eta_{sp,2}|^2}{\delta-\hbar\omega}\nonumber \\
& +\frac{|\eta_{sp,2}|^2}{\delta-5\hbar \omega},\\
\tilde{\varepsilon}_p =\ & \varepsilon_{p,0}-3\hbar\omega - \frac{|\eta_{sp,1}|^2}{\delta-2\hbar \omega}-\frac{|\eta_{sp,1}|^2}{\delta-5\hbar\omega}-\frac{|\eta_{sp,2}|^2}{\delta-\hbar\omega}\nonumber \\
& -\frac{|\eta_{sp,2}|^2}{\delta-5\hbar \omega}.
\end{align}
The effective coupling term amounts to
\begin{align}
\tilde{\eta}_{sp} =\ & \eta_{sp,3} + \frac{\eta_{sp,1}(\varepsilon_{p,2}-\varepsilon_{s,2})}{2}\left(\frac{1}{\delta-2\hbar \omega}-\frac{1}{2\hbar \omega}\right)\nonumber\\
& +\frac{\eta_{sp,2}(\varepsilon_{p,1}-\varepsilon_{s,1})}{2}\left(\frac{1}{\delta-\hbar \omega}-\frac{1}{\hbar \omega}\right)\nonumber  \\
&+\frac{\eta_{sp,1}(\varepsilon_{p,1}^2-\varepsilon_{s,1}^2)}{4(\hbar \omega)^2}\left(\frac{2(\hbar\omega)^2}{(\delta-2\hbar\omega)(\delta-\hbar \omega)}+1\right)\nonumber\\
&+\frac{\eta_{sp,1}\varepsilon_{s,1}\varepsilon_{p,1}}{2 \hbar \omega \left(\delta-\hbar\omega\right)}+\frac{\eta_{sp,1}^3}{2\hbar\omega \left(\delta-2\hbar\omega\right)},
\end{align}
with the three-photon detuning $\delta=\varepsilon_{s,0}-\varepsilon_{p,0}+3\hbar\omega$.

For the three configurations, we plot in Fig.~\ref{figA3} the spectrum of the effective Hamiltonians (solid lines) on top of the results from a Floquet-Bloch band calculation (dots in light blue).
We take up to fifth order terms into account for the tight-binding expansion ($P=5$) used to calculate the matrix elements from Table~\ref{tab:JacobiAnger}.
In Fig.~\ref{figA3}(a) the single photon coupling is shown.
The gap opens around a quasimomentum value $q \approx 0.25\ \pi/a$ between the two lowest bands and is well captured by the effective theory.
Since we only take into account two bands, the second appearing gap at $q \approx 0.5\ \pi/a$ is not captured.
In addition, the third band is inducing strong energy shifts around the band center, especially for the upper effective band.
A more precise model is gained by expanding the effective Hamiltonian to a three band system.

In the case of the three-photon resonance, see Fig.~\ref{figA3}(b), we compare effective theory and Floquet-Bloch calculation for the single-frequency driving as well as two-frequency driving with additional third harmonic.
The parameters are chosen such to arrive at the gap closing transition measured in the main text.

In the situation of a two-photon resonance, see Fig.~\ref{figA3}(c), the effective spectra are benchmarked in the three situations of single frequency driving, two-frequency driving at the critical values for right as well as left gap closing transition. \\
The Floquet-Bloch data includes the third band which hybridizes quite strongly with the second band but does not disturb the effective lowest band.
Therefore, this method is very well suited to engineer a single band model that is defined by the lowest band.

\begin{figure}
    \includegraphics{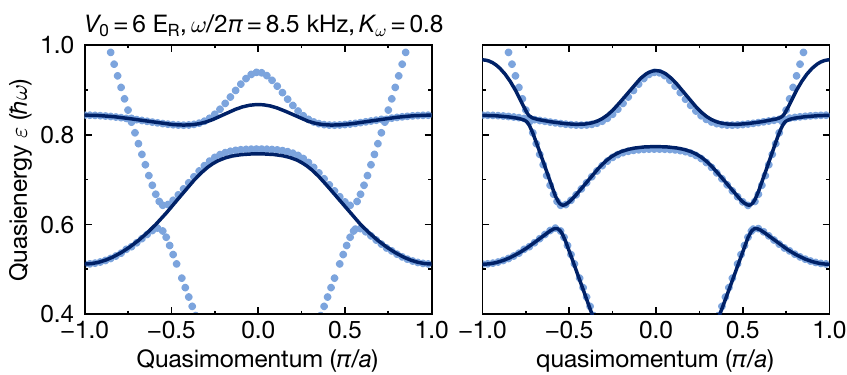}
    \caption{Compare effective two- and three-band models for single photon resonance ($\varepsilon_{p,0}-\varepsilon_{s,0} \simeq \hbar\omega$). The points displayed on the plots corresponds to the Floquet-Bloch calculation. The solid lines is the dispersion for the two-band model (left) and the three band model (right).\label{figA4}}
\end{figure}
Expanding the effective theory to three bands makes it possible to get a more precise model in the single photon resonance case.
The effective Hamiltonian matrix is extended to
\begin{equation}
\tilde{H} = 
\begin{pmatrix}
\tilde{\varepsilon}_{s} & \tilde{\eta}_{sp} & \tilde{\eta}_{sd} \\
\tilde{\eta}_{sp}^\ast & \tilde{\varepsilon}_{p} & \tilde{\eta}_{pd} \\
\tilde{\eta}_{sd}^\ast & \tilde{\eta}_{pd}^\ast & \tilde{\varepsilon}_d 
\end{pmatrix},
\label{eqn:Heff3}
\end{equation}
with the diagonal elements
\begin{align}
\tilde{\varepsilon}_s =\ & \varepsilon_{s,0} +\frac{|\eta_{sp,1}|^2}{\delta_{sp}-2\hbar \omega}+\frac{|\eta_{sd,1}|^2}{\delta_{sd}-\hbar \omega}+\frac{|\eta_{sd,1}|^2}{\delta_{sd}-3\hbar \omega}, \\
\tilde{\varepsilon}_p =\ & \varepsilon_{p,0}-\hbar\omega - \frac{|\eta_{sp,1}|^2}{\delta_{sp}-2\hbar \omega}+ \frac{|\eta_{pd,1}|^2}{\delta_{pd}-2\hbar \omega}+\frac{|\eta_{pd,2}|^2}{\delta_{pd}-3\hbar\omega}\nonumber\\
&+\frac{|\eta_{pd,2}|^2}{\delta_{pd}+\hbar\omega}+\frac{|\eta_{pd,3}|^2}{\delta_{pd}+2\hbar \omega}+\frac{|\eta_{pd,3}|^2}{\delta_{pd}-5\hbar\omega}, \\
\tilde{\varepsilon}_d =\ & \varepsilon_{d,0}-2\hbar\omega - \frac{|\eta_{sd,1}|^2}{\delta_{sd}-\hbar \omega}-\frac{|\eta_{sd,1}|^2}{\delta_{sd}-3\hbar \omega}- \frac{|\eta_{pd,1}|^2}{\delta_{pd}-2\hbar \omega}\nonumber\\
&-\frac{|\eta_{pd,2}|^2}{\delta_{pd}-3\hbar\omega}-\frac{|\eta_{pd,2}|^2}{\delta_{pd}+\hbar\omega}-\frac{|\eta_{pd,3}|^2}{\delta_{pd}+2\hbar \omega} \nonumber\\
&-\frac{|\eta_{pd,3}|^2}{\delta_{pd}-5\hbar\omega}.
\end{align}
If we consider the coupling paths between the states, we obtain the off-diagonal elements
\begin{align}
\tilde{\eta}_{sp} =\ & \eta_{sp,1} + \frac{\eta_{sp,2}(\varepsilon_{p,1}^\ast-\varepsilon_{s,1}^\ast)}{2}\left(\frac{1}{\delta_{sp}+\hbar \omega}+\frac{1}{\hbar \omega}\right)\nonumber\\
& + \frac{\eta_{sp,1}^\ast(\varepsilon_{p,2}-\varepsilon_{s,2})}{2}\left(\frac{1}{\delta_{sp}-2\hbar\omega}+\frac{1}{2\hbar\omega}\right),\\
\tilde{\eta}_{sd} =\ & \eta_{sd,2} + \frac{\eta_{sd,1}(\varepsilon_{d,1}-\varepsilon_{s,1})}{2}\left(\frac{1}{\delta_{sd}-\hbar \omega}-\frac{1}{\hbar \omega}\right)\nonumber\\
& + \frac{\eta_{sp,3}\eta_{pd,1}^\ast}{2}\left(\frac{1}{\delta_{sp}+\hbar\omega}-\frac{1}{\delta_{pd}-\hbar\omega}\right)\nonumber\\
& + \frac{\eta_{pd,3}\eta_{sp,1}^\ast}{2}\left(\frac{1}{\delta_{sp}-\hbar\omega}-\frac{1}{\delta_{pd}+\hbar\omega}\right), \\
\tilde{\eta}_{pd} =\ & \eta_{pd,1} +  \frac{\eta_{pd,2}^\ast \varepsilon_{d,1}}{2} \left(\frac{1}{\delta_{pd}+\hbar\omega}+\frac{1}{\hbar\omega}\right) \nonumber  \\
& + \frac{\eta_{pd,1}^\ast \varepsilon_{d,2}}{2} \left(\frac{1}{\delta_{pd}-2\hbar\omega}-\frac{1}{2\hbar\omega}\right),
\end{align}
where the detunings are $\delta_{sp} = \varepsilon_{s,0}-\varepsilon_{p,0}+\hbar\omega$, $\delta_{sd} = \varepsilon_{s,0}-\varepsilon_{d,0}+2\hbar\omega$
and $\delta_{pd} = \varepsilon_{p,0}-\varepsilon_{d,0}+\hbar\omega$.
The three band model is in very good agreement to the exact spectrum as shown in Fig.~\ref{figA4} on the left in direct comparison to the two-band model on the right.
Only for the third band at the band edges we can see a clear difference.
It would be necessary to include the next higher bands to correct for this deviation.
However, at the chosen lattice depth of $V_X = 6~\text{E}_\text{R}$ it is very inconvenient to use a tight-binding approximation and perturbation calculation becomes unpractical.

\subsubsection{Extracting gaps}
\label{sec:EffDisp}
The effective models can be used to extract the gap for various driving parameters.
In the case of a two-band model the Hamiltonian can be written in Bloch form 
\begin{equation}
\tilde{H} = E(q)\mathbb{1} - \boldsymbol{h}(q)\cdot \boldsymbol{\sigma}, 
\label{eqn:Hbloch}
\end{equation}
where $\boldsymbol{\sigma}=(\sigma_x,\sigma_y,\sigma_z)^T$ is the Pauli matrix vector.
The eigenenergies are
\begin{equation}
\varepsilon_{\pm} = E(q)\pm \sqrt{h_x^2+h_y^2+h_z^2},
\end{equation}
and the gap is simply proportional to the second term
\begin{equation}
\Delta = 2\sqrt{h_x^2+h_y^2+h_z^2} = 2 |\boldsymbol{h}|.
\end{equation}
The theory plots in Fig.~\ref{fig3} and Fig.~\ref{fig4} of the main text are calculated using the effective theory.

Since the effective theory captures the Floquet-Bloch band structure very well, we also use it to estimate the gradients of the dispersion which are used to calculate the transition speed for the Landau-Zener formula given in \cite{Suppl}.
In the case of the single photon resonance we use the three band model to obtain higher precision in this gradient.

\clearpage

\newcommand{\Erec}{E_{\text{rec}}}
\newcommand{\klat}{k_{L}}

\makeatletter
\setcounter{section}{0}
\setcounter{subsection}{0}
\setcounter{figure}{0}
\setcounter{equation}{0}
\renewcommand{\bibnumfmt}[1]{[S#1]}
\renewcommand{\citenumfont}[1]{S#1}
\renewcommand{\thefigure}{S\@arabic\c@figure}
\renewcommand{\theequation}{S\@arabic\c@equation}
\makeatother

\centerline{\Large \textbf{Supplemental material}}

\section{Experimental methods}

The experiment starts with a gas of $^{87}$Rb bosonic atoms in the sublevel $m_F=2$ of the $F=2$ manifold, which is trapped in a harmonic optical dipole trap. The atoms are evaporatively cooled down to Bose-Einstein condensate at the background scattering length. The atom number is calibrated with strong saturation absorption imaging technique \cite{Reinaudi2007}. We then ramp up a magnetic gradient to counteract gravity and ramp down the dipole trap at the same time. The dipole trap is further ramped to zero when we ramp up the optical lattice. Before loading the lattice we have a mean number of atoms of $15(2)\times 10^{3}$ with a condensate fraction of $44(6)\%$.

The one-dimensional optical lattice consists of a retro-reflected laser beam of wavelength $\lambda = 1064 \, \text{nm}$. The lattice potential seen by the atoms is
\begin{equation} V(x) = V_X \cos^2(k_{L}x) , 
\label{eqn:LatticePotential}
\end{equation}
with $k_{L}=2\pi/\lambda$. The lattice depths $V_X$ is measured in units of the recoil energy $E_R=h^2/2M\lambda^2$ ($h$ is the Planck constant and $M$ the mass of the Rubidium atoms). The lattice depth is calibrated using amplitude modulation on a $^{87}$Rb Bose-Einstein condensate. There is also a very shallow lattice along z-direction to trap the atoms against a residual gradient along the $y$-direction. The parameters of our lattice configuration are shown in Table \ref{Table_LatticeConfiguration}. The Hubbard parameters $t$ is numerically calculated from the Wannier functions of the lattice potential, which we obtain from band-projected position operators \cite{Uehlinger2013}. Our red-detuned lattice also induces an external confinement and the corresponding trap frequency is also shown in Table \ref{Table_LatticeConfiguration}.

\subsection{Periodic driving}

The periodic driving is realized with a piezo-electric actuator which modulates the position of the retro-reflecting mirror for the $X$ lattice beam at a frequency $\omega/2\pi$ and displacement amplitude $\Delta L$. 
The phase of the retro-reflected $X$ lattice beam is therefore shifted with respect to the incoming one such that the time-modulated ($\tau$) lattice potential can be expressed as $V(x,\tau)=V(x-x_0(\tau))$. 
For a two-frequency driving scheme we use the waveform
\begin{equation}
x_0(\tau)=\Delta L_\omega \cos(\omega \tau) + \Delta L_{l\omega} \cos(l\omega\tau+\varphi),
\label{eqn:displacement}
\end{equation}
where $l$ denotes the order of the higher harmonic contribution that is used and $\varphi$ the relative phase to its fundamental counterpart. The length displacements $\Delta L_{l\omega}$ are associated with the dimensionless amplitude via 
\begin{equation}
K_{l\omega}=M\Delta L_{l\omega} l \omega a/\hbar,
\label{eqn:K}
\end{equation}
where $a$ is the lattice constant along the $x$-direction ($\hbar=h/2\pi$). 
The amplitude and phase of the mirror displacement is calibrated by measuring the phase modulation caused by the periodic driving using a Michelson interferometer.
The previously used design of the actuator-mirror configuration in \cite{Gorg2019} has been updated. 
The first mechanical resonance of the actuator-mirror configuration is pushed to high frequencies ($\sim 60\,\mathrm{kHz}$) by using a single-stack, piezo-electric actuator (Noliac NAC2013) combined with a tungsten mount ($216\,\mathrm{g}$) and a quarter-inch mirror ($3\,\mathrm{mm}$ thick).
The residual frequency and phase dependence is caused by the capacitive load ($\sim 190\,\mathrm{nF}$) of the piezo-electric actuator driven via a voltage amplifier (PiezoDrive PX200) and shows a smooth behavior that is calibrated out via the above mentioned method.
The systematic error due to this calibration method amounts to $0.5 \%$ of the driving strength and $0.25^\circ$ on the relative phase.
Furthermore, we acquire a statistical error on the strength and phase of the same amount.
Since the phase calibration method only works reliable for strengths as low as $K_\omega=0.5$, we extrapolate the calibration values for lower driving strengths.
We detect a systematic phase shift for the optimal phase for a gap closing in a three-photon resonance correlated to low values of $K_{3\omega}<0.3$.
In principle, we can reach with this system a bandwidth of 100~kHz with driving strengths up to $K_\omega=3.5$ for $\omega/2\pi \geqslant 2$~kHz.

\begin{table}[bt]
\begin{tabular}{ c  c }
\hline\hline
parameter & value \\
\hline
$V_{X,Z} \left(E_R\right)$ & 6.0(1),0.82(2) \\
$t_{x,z}/h \left(\text{Hz}\right)$ & 224(6),812(3) \\
$f^{trap}_{x,y,z}/h \left(\text{Hz}\right)$ & 7.4(3),23(2),21(3) \\
\hline\hline
\end{tabular}
\caption{Parameters of the lattice used in this experiment. 
Errors in the lattice depths account for an uncertainty of the lattice calibration and an additional statistical error due to fluctuations of the lattice depth.
The value and error on the tunneling rates $t_{x,z}$ result from the uncertainty of the lattice depth. The trap frequency is measured by kicking the atoms using magnetic gradient without retro-reflected beam.
}
	\label{Table_LatticeConfiguration}
\end{table}

\subsection{Bloch Oscillation}
\label{sec:BO}
The Bloch oscillation used to detect the gap is induced by a magnetic gradient which is calibrated by measuring its frequency. The center and the size of the Brillouin Zone is measured with Bragg diffraction where we flash the lattice and extract the position of the $2\hbar k_{L}$ diffraction peaks.\\

The frequency of the Bloch oscillation $\nu_{\text{BO}}$ gives the energy resolution with which we can probe the Floquet-Bloch gaps. In our setup the resolution is limited by the minimal trapping frequency in the direction of the Bloch oscillation which we can achieve without untrapping the atoms. If we use the levitation scheme described in the previous section we can achieve the minimal trapping frequency in $x$-direction stated in Table~\ref{Table_LatticeConfiguration}. It is given by the confinement of the orthogonal lattice beam. On the other hand, at large $K_\omega$ some other gaps (multi-photon resonance to higher bands) which we do not want to probe become non-negligible. In that case we increase the frequency of the Bloch oscillation so that the detection is only sensitive to the largest gap that we are interested in.\\

The value of the Floquet-Bloch gap $E_{\text{gap}}$ is calculated from measured transition rates $P_{\text{trans}}$ with the Landau-Zener formula
\begin{equation}
P_{\text{trans}} = 1-\exp\left(-\pi^{2}\frac{E_{\text{gap}}^{2}}{\Delta\nu/\Delta \tau}\right),
\label{LZ}
\end{equation}
where $E_{\text{gap}}$ is in units of Hz and $\Delta\nu/\Delta \tau$ is the energy sweep rate in units of Hz$^{2}$.
The energy sweep rate is calculated from the frequency of Bloch oscillation $\nu_{\text{BO}}$ and the effective dispersion relations of s- and p-band $\tilde{\varepsilon}_{s,p}(q)$ using a two-band approximation
\begin{equation}
\frac{\Delta\nu}{\Delta \tau} = \frac{\nu_{\text{BO}}}{\hbar}\frac{\partial}{\partial q}\left[\tilde{\varepsilon}_p(q)-\tilde{\varepsilon}_s(q)\right]_{q=q_{\text{gap}}}.
\end{equation}
For details on the calculation of the effective dispersion relation see the Appendix of the main text.

\subsection{Detection methods}

The transferred fraction is obtained from the band-mapping detection, where we ramp down the optical lattice slowly (1 ms) after the modulation such that the atoms stay adiabatically in their band and the quasi-momentum ($q$) is mapped to real momentum. After that we switch off the magnetic levitation and allow for 25 ms time of flight (TOF) to map momentum onto position and then take an absorption image. To determine the transferred fraction we fit two Gaussian functions to the two clouds which correspond to the transferred and not-transferred part and capture the atom number for each cloud from the fitting.

\section{Floquet Bloch bandstructure of a shaken optical lattice}
\label{sec:FB}

The single-particle spectrum of a periodically shaken optical lattice with translational symmetry is derived via Floquet's theorem and the Trotter decomposition.
In this derivation we closely follow \cite{Holthaus2015}.

\subsection{Static bandstructure of an optical lattice}

The spectrum of a single particle in a static cosine-lattice (see Eq.~\ref{eqn:LatticePotential}) can be obtained by numerically solving the eigenvalue problem
\begin{align}\label{eqn:eigenvaluestatic}
  & \left(-\frac{\diff^2}{\diff z^2} - 2i \frac{q}{\klat} \frac{\diff}{\diff z} + \left(\frac{q}{\klat}\right)^2 + \frac{V_X}{2\Erec}\cos(2z) \right) u_q^n(z)  \nonumber\\
  &= \frac{E(q)}{\Erec} u_q^n(z) 
\end{align}
for the periodic Bloch functions $u_q^n(z)$ at quasimomentum $q$ with band index $n$.
The periodicity of the lattice is $a = \lambda/2 = \pi/\klat$ and $V_X = 6.0\ \Erec$ is the lattice depth.
Equation~\ref{eqn:eigenvaluestatic} has been made dimensionless by scaling energy in units of recoil $\Erec = \frac{(\hbar \klat)^2}{2M}$ with $m$ being the mass of a $^{87}$Rb atom and by introducing the dimensionless coordinate $z = \klat x$.
The operators $-\frac{\diff^2}{\diff z^2}$, $\frac{\diff}{\diff z}$, and $\cos(2z)$ can be written as matrices in the basis of $\pi$-periodic functions \cite{Holthaus2015}.
We typically truncate the lattice Hamiltonian to $15\times15$ entries.

\subsection{Floquet-Bloch bandstructure}

The Floquet drive is realized by sinusoidally modulating the position of the retro-reflecting mirror (see Eqn.~\ref{eqn:displacement}) that creates the optical standing wave.
Typical values of $\Delta L_\omega$ in this work are on the order of $0.01\ a - 0.3\ a$.
In order to incorporate the periodic drive into the lattice eigenvalue problem (Eq.~\ref{eqn:eigenvaluestatic}) it is most convenient to work in a frame rotating with the modulated position $x$.
This can be achieved by applying a unitary transformation which yields a time-dependent `vector potential' that is added to the momentum operator in the Hamiltonian
\begin{equation}
H_{\text{rot}}(\tau) = \frac{[\hat{p} - A(\tau)]^2}{2M} + V(\hat{x})~.
\end{equation}
Sometimes this frame of reference is also referred to as the `rotating frame'.
For the `vector potential' $A(\tau)$ we then have
\begin{equation}
  \label{eqn:xm-rot}A(\tau) = M\dot{x}_0(\tau) = -\frac{\hbar}{a}\left[ K_\omega\sin(\omega\tau)+K_{l\omega}\sin(l\omega\tau+\varphi)\right].
\end{equation}
The dimensionless driving strengths $K_\omega, K_{l\omega}$ defined in Eqn.~\ref{eqn:K} will be convenient in the calculation following below.

Since the resulting time-dependent Hamiltonian
\begin{align}
H_{\text{rot}}(\tau) =& -\frac{\diff^2}{\diff z^2} - 2i \left[\frac{q}{\klat} + \frac{K_\omega}{\pi} \sin(\omega\tau)\right.\nonumber\\
&\left. + \frac{K_{l\omega}}{\pi} \sin(l\omega\tau+\varphi)\right] \frac{\diff}{\diff z} + \left(\frac{q}{\klat}\right)^2\nonumber \\
& + \frac{V_X}{2\Erec}\cos(2z)
\end{align}
is periodic both in time and in space, we can apply Floquet's theorem and find solutions as spatio-temporal Bloch waves.
The energy shift resulting from the square of the vector potential is rotated away.
We obtain the time-evolution operator $\mathcal{U}(\tau_0 + T,\tau_0)$ over one driving period $T = \omega/(2\pi)$ via the Trotter decomposition
\begin{align}
\mathcal{U}(\tau_0 + T, \tau_0) &= \mathcal{T} \exp\left[-\frac{i}{\hbar}\int_{\tau_0}^{\tau_0 + T} H_{\text{rot}}(\tau) \diff \tau \right]\\
  &\simeq \exp\left[-\frac{i}{\hbar}\sum_{j = 0}^{N-1} H_{\text{rot}}(\tau_j) \Delta\tau \right]\\
  \nonumber  &= \prod_{j = 0}^{N-1} \exp\left[-\frac{i}{\hbar} H_{\text{rot}}(\tau_j) \Delta\tau\right] + \mathcal{O}\left(\Delta\tau^2\right)
\end{align}
where $\mathcal{T}$ denotes time-ordering.
The driving period $[\tau_0, \tau_0 + T[$ is discretized in $N$ steps as $\tau_j = \tau_0 + j \Delta\tau$ with $\Delta \tau = T/N$.
For typical driving strengths of $K_\omega \lesssim 1$ a discretization into $N = 50$ is sufficient; obtaining faithful results for larger driving strengths requires a finer discretization.
Alternatively, the time-evolution operator $\mathcal{U}(T,0)$ can be obtained by directly integrating the time-dependent Schrödinger equation.
However, we find that for our purposes the Trotter decomposition is more efficient.
Since we are only interested in the quasienergy spectrum, we can take the Floquet gauge $\tau_0 = 0$ without loss of generality.  

The resulting quasienergies $\varepsilon_n(q)$ are encapsulated in the Floquet multipliers $\{e^{-i\varepsilon_n(q)T/\hbar}\}$ which are the eigenvalues of $\mathcal{U}(T, 0)$.
The quasienergies form the Floquet-Bloch bandstructure which describes the exact spectrum of a single particle in a homogeneous, shaken optical lattice, limited only by the numerical discretization and the truncation of the Hilbert space.
This description includes all transitions to and within higher bands, as well as any additional non-perturbative effects beyond the usual high-frequency regime (rotating-wave approximation, high-frequency expansion, Magnus expansion).

\subsection{Numerical evaluation of the Floquet-Bloch gaps}

Single- and multi-photon resonances between Bloch bands lead to gap openings in the Floquet-Bloch quasienergy spectrum.
In order to numerically evaluate the size of these gaps the evaluated Floquet states (eigenstates of $\mathcal{U}(T,0)$) are sorted according to their overlap with the static Bloch bands.
At the quasimomentum value where an interband coupling occurs the order of this sorting is changed and we can extract the size of the gap. 
For large driving strengths the single-gap picture breaks down and additional resonances appear.
In the case of the single-photon resonance, we can reliably extract the gap until $K_\omega = 1.0$. For the two- and three-photon resonances, we can evaluate the gap until $K_\omega = 1.25$ and $K_\omega = 1.6$, respectively.
For the largest values of shaking strength, specifically for the three-photon resonances above $K_\omega>0.68$, we linearly increase the shaking frequency in order to keep the resonance roughly fixed at a specific quasimomentum, thereby counteracting the AC-Stark shift.

For the computation of the single-frequency gap openings, we sample quasimomentum between 0 and $\pi/a$ in 501 steps.
Doubling the $q$-sampling does not change the absolute gap values by more than 3~Hz.
At large driving strengths the admixture of higher (static) bands can lead to `outliers' in the maximum gap values. We ignore these in the calculations for Fig.~2 of the main text.
For the computation of the two-frequency gap closings, we sample quasimomentum between 0 and $\pi/a$ in 101 steps.

\end{document}